\documentclass[preprint]{aastex631}

\usepackage{amsmath}
\usepackage{multirow}

\newcommand{\bma}[1]{\boldsymbol{#1}}
\newcommand{\unit}[1]{\ensuremath{\,\mathrm{#1}}}

\usepackage[normalem]{ulem}

\shorttitle{ffHD II}
\shortauthors{Zhou et al.}
\graphicspath{{./}{figures/}}

\begin{document}
\title{Frozen-field Modeling of Coronal Condensations with MPI-AMRVAC \\ II: Optimization and application in three-dimensional models}
\correspondingauthor{Yuhao Zhou}
\email{yuhao.zhou@kuleuven.be}

\author[0000-0002-4391-393X]{Yuhao Zhou}
\affiliation{Centre for mathematical Plasma-Astrophysics (CmPA), KU Leuven \\
Celestijnenlaan 200B, 3001 Leuven, Belgium}

\author[0000-0001-8164-5633]{Xiaohong Li}
\affiliation{Max Planck Institute for Solar System Research \\ 
Göttingen D-37077, Germany}
\affiliation{Centre for mathematical Plasma-Astrophysics (CmPA), KU Leuven \\
Celestijnenlaan 200B, 3001 Leuven, Belgium}

\author[0000-0002-8975-812X]{Jack M. Jenkins}
\affiliation{Centre for mathematical Plasma-Astrophysics (CmPA), KU Leuven \\
Celestijnenlaan 200B, 3001 Leuven, Belgium}

\author[0000-0002-8002-7785]{Jie Hong}
\affiliation{Institute for Solar Physics, Dept. of Astronomy, Stockholm University, AlbaNova University Centre, SE-106 91 Stockholm, Sweden}

\author[0000-0003-3544-2733]{Rony Keppens}
\affiliation{Centre for mathematical Plasma-Astrophysics (CmPA), KU Leuven \\
Celestijnenlaan 200B, 3001 Leuven, Belgium}

\begin{abstract}
The frozen-field hydrodynamic (ffHD) model is a simplification of the full magnetohydrodynamical (MHD) equations under the assumption of a rigid magnetic field, which significantly reduces computational complexity and enhances efficiency.
In this work, we combine the ffHD prescription with hyperbolic thermal conduction (TC) and the Transition Region Adaptive Conduction (TRAC) method to achieve further optimization. 
A series of two-dimensional tests are done to evaluate the performance of the hyperbolic TC and the TRAC method. 
The results indicate that hyperbolic TC, while showing limiter-affected numerical dissipation, delivers outcomes comparable to classic parabolic TC. 
The TRAC method effectively compensates for the underestimation of enthalpy flux in low-resolution simulations, as evaluated on tests that demonstrate prominence formation.
We present an application of the ffHD model that forms a three-dimensional prominence embedded in a magnetic flux rope, which develops into a stable slab-like filament. 
The simulation reveals a prominence with an elongated spine and a width consistent with observations, highlighting the potential of the ffHD model in capturing the dynamics of solar prominences. 
Forward modeling of the simulation data produces synthetic images at various wavelengths, providing insights into the appearance of prominences and filaments in different observational contexts.
The ffHD model, with its computational efficiency and the demonstrated capability to simulate complex solar phenomena, offers a valuable tool for solar physicists, and is implemented in the open-source MPI-AMRVAC framework. 
\end{abstract}


\section{Introduction}
\label{sec1}
In research on solar physical activities, the magnetohydrodynamical (MHD) equations are often employed for describing large-scale phenomena. 
Through analytical derivations or numerical simulations, MHD equations successfully explain many physical phenomena on the Sun, as noted in modern textbooks {\citep{prie2014, goed2019}}.
Over the past several decades, numerous numerical simulation tools based on MHD equations have been developed, such as MPI-AMRVAC \citep{kepp2023}, Athena++ \citep{ston2020}, Bifrost \citep{gudi2011}, FLASH \citep{fryx2000}, LAREXD \citep{arbe2001}, MANCHA \citep{mode2024}, MURaM \citep{remp2017}, Pencil \citep{penc2021}, PLUTO \citep{mign2012}, etc.
These tools collectively succeeded in simulating large-scale solar and astrophysical phenomena with unprecedented agreement between simulations and observations.

However, solving these equations efficiently remains challenging due to the complexity of the nonlinear, anisotropic MHD equations, especially when considering physical factors in addition to ideal MHD behavior. 
These physical factors include, but are not limited to, radiative transfer, radiative losses, anisotropic thermal conduction (TC), ambipolar diffusion, Hall effects, Joule dissipation, etc. 
Given limited computational resources, maintaining stability in numerical simulations at finite resolutions while retaining accuracy also presents a modern challenge. 

For solar coronal applications, the MHD equations may be simplified to enhance numerical stability and computational efficiency. Since the plasma $\beta$ parameter (ratio of thermal pressure to magnetic pressure) in the solar corona is relatively low, the dynamics are dominated by the magnetic field, while the fluid motion exerts little feedback on the magnetic field. 
If the time scale of the phenomenon under study is shorter than the evolutionary time scale of the magnetic field, the magnetic field can be considered as a rigid body that guides all motion. 
In such cases, the MHD description could be simplified to pure hydrodynamic (HD) behavior along a field line. 
This approach is often used in studies of flares and coronal loops, and employed in one-dimensional (1D) codes like {HYDRAD \citep{brad2003}}, FLARIX \citep{vara2010} or RADYN \citep{allr2015, carl2023}, as well as in studies of prominence formation \citep[e.g.,][]{anti1999, karp2005, xia2011, guo2022} or their oscillations \citep[e.g.,][]{zhan2012, zhou2017}. 
These studies have been purely 1D \citep[even in the pseudo-three-dimensional settings in][]{guo2022}, but recently two-dimensional (2D) generalizations that include actual MHD feedback processes in arcade configurations have emerged in \citet{zhou2020, jerc2022, jerc2023}. 
{
In these 2D works, the arcade configurations have an invariant direction and allow for arbitrary field line shapes by extending a chosen 1D field line along the invariant direction. 
However, this approach cannot be utilized when studying three-dimensional (3D) flux ropes, as it is challenging to find realistic and stable magnetic fields that correspond to such 2D models.
}
Here, we rather focus on a different generalization to multi-dimensional settings, that does not include any MHD feedback, as pioneered by \citet{mok2005}.
\citet{mok2005} applied it to study 3D active region evolution, where the authors ran 3D HD simulations while restricting the direction of the flows to be field-aligned.
Subsequent works applied this method to studies of heating models, thermodynamic evolution and forward modeling of coronal loops \citep{mok2008, lion2013, wine2014, wine2016, mok2016}.

In our previous work \citep{zhou2024}, we presented the governing equations in conservative form for this simplified MHD model. 
Due to the fixed nature of the magnetic field lines in the model, we named it frozen-field hydrodynamic (ffHD) model. 
We implemented the ffHD model in the open-source MPI-AMRVAC code\footnote{\url{http://amrvac.org}} and applied it to 2D simulations of evaporation-condensation prominence formation in magnetic arcades, proving its effectiveness. 
We also compared this model with the MHD model and a pseudo-2D model composed of hundreds of 1D simulations (on independently treated field lines), analyzing the differences among them.
While this comparison proved insightful, the arcade structure of the magnetic field in these simulations was relatively simple, and computational resources required for instead doing full high-resolution MHD simulations are indeed affordable. 
This made the advantages of using ffHD for 2D settings not really pronounced. 

In 3D scenarios, many prominence-related studies favor adopting magnetic flux rope configurations because statistically, most prominences are supported by magnetic flux ropes rather than shear arcades \citep{ouya2017}. 
Flux ropes are more complex than arcades, and MHD simulations involving them face more numerical challenges, especially since the inherent 3D structure may need to be studied with insufficient resolution and hence may suffer from numerical dissipation. 
\citet{xia2016} conducted the first 3D MHD numerical simulation of prominence formation through the evaporation--condensation mechanism (improving on an earlier 3D study by \citet{xia2014} where no true fine-structure was realized).
The simulation showed a rather fragmented and very dynamic prominence structure, which is different from the commonly used simple slab models for prominences \citep[e.g.,][]{kipp1957}, and it did not yet fully agree with observational filamentary (thread-like) prominences \citep[e.g.,][]{lin2005}.
In recent high-resolution simulations of prominence formation \citep{zhou2020, jenk2022, donn2024}, clearer elongated structures along the spine of the prominence can be observed, which {demonstrates} the importance of high resolution in simulations of prominence formation. 
Thus, the ffHD model may find greater applicability in 3D scenarios.

To conduct 3D simulations more efficiently, several optimizations can be introduced. 
First and foremost is the need to incorporate anisotropic TC, and numerically handle the parabolic (and possibly stiff) TC term. 
\citet{xia2018} reported on the implementation strategy for anisotropic TC in MPI-AMRVAC, which uses properly slope-limited discretizations and can benefit from using adaptive mesh refinement (AMR). 
Still, TC in 3D MHD settings can severely slow computations, even when using the Super Time-Stepping (STS) method \citep{meye2012} in the operator-split evaluation of the TC term. 
The current version of MPI-AMRVAC \citep{kepp2023} includes multiple STS variants that are compatible with AMR and help alleviate {the more restrictive Courant condition associated with parabolic terms, which scales with the square of the grid spacing. 
However, the proportion of computational time devoted to parabolic TC remains significant.}
In the ffHD module, where only three equations are involved, the relative computational time consumed by accurately calculating the parabolic form of TC becomes even more pronounced.
Therefore, optimizing the computation of the TC term is a priority in this model. 
A viable option is the hyperbolic TC used in the MuRAM code \citep{remp2017}, and also exploited in Pencil \citep{warn2020} and MANCHA \citep{nava2022}. 
By rewriting traditional parabolic TC to hyperbolic, the TC flux also gains a finite propagation speed, thereby no longer being strictly limited by the time step of parabolic TC, which can significantly enhance computational speed. 
In this paper, we will also cross-validate TC treatments in MPI-AMRVAC, as of now extended with a hyperbolic approach.

Another point of optimization for TC concerns the correction of enthalpy fluxes. 
It has been noted that in simulations where resolution is insufficient to fully resolve the chromosphere--corona transition region, the enthalpy flux causing evaporation is generally underestimated, leading to a coronal density lower than that in high-resolution simulations \citep{brad2013}. 
The required resolution threshold could be less than 1 km in some cases \citep{john2019a, john2019b}, which is clearly unfeasible for large-scale studies in 3D unrefined mesh settings. 
For large-scale 3D simulations, it is challenging to meet this threshold even with AMR capabilities, so certain corrections are necessary to compensate for the underestimated enthalpy flux. 
This correction is particularly crucial in evaporation-condensation models, as shown e.g. in our previous work \citep{zhou2021}. 
A commonly used and effective method is the transition region adaptive conduction (TRAC) method \citep{john2019a}, which is associated with a minimal broadening of the transition region and produces the smallest errors in thermal dynamics \citep{hows2023}.
The multi-dimensional TRAC method has been implemented in a number of variants within MPI-AMRVAC \citep{zhou2021, kepp2023}, and a recent update that also included parallelization of the magnetic field line tracing module (which is used in some of the non-local TRAC variants) led to further improved efficiency. 
Here, we demonstrate the combination of TC variants, usage of TRAC within the ffHD module on relevant evaporation-condensation tests, and on a full 3D flux rope application exploiting AMR.

The structure of this paper is as follows: Sect.~\ref{sec2} will introduce various optimizations, enhancements, and efficiency tests of the ffHD module.
{
In Sect.~\ref{sec3}, we introduce the full 3D setup of our ffHD simulations of prominence formation in a magnetic flux rope configuration involving the evaporation-condensation mechanism, followed by a detailed analysis of the simulation results.
}
Finally, conclusions and discussions will be presented in Sect.~\ref{sec4}.

\section{Further Optimizations for ffHD}
\label{sec2}
\subsection{Computational Considerations in ffHD}
\label{sec21}
The adiabatic ffHD equations with frozen-field aligned gravity as derived in \citet{zhou2024} are
\begin{eqnarray}
    \frac{\partial \rho}{\partial t}+\nabla \cdot  \left (  \rho v_{\parallel} \bma{\hat{b}}\right ) &=& 0,
    \label{eq21}\\
    \frac{\partial \left(\rho v_{\parallel}\right)}{\partial t}+\nabla \cdot \left [  \left (  \rho v^2_{\parallel} +p\right ){\bma{\hat{b}}}\right ] &=& \rho g_{\parallel} + p (\nabla\cdot \bma{\hat{b}}),
    \label{eq22}\\
    \frac{\partial E}{\partial t} + \nabla \cdot \left[\left(E + p\right) v_{\parallel} \bma{\hat{b}}\right] &=& \rho g_{\parallel} v_{\parallel}.
    \label{eq23}
\end{eqnarray}
Here, $v_{\parallel}$ and $\rho g_{\parallel}$ are field-aligned velocity and gravity, respectively. 
$\bma{\hat{b}} = (b_x, b_y, b_z)$ is the unit magnetic field vector along the field line.
Hydrodynamic energy density relates to pressure $p$ and density $\rho$ as in $E=p/\left(\gamma-1\right) + \rho v_{\parallel}^2/2$, where $\gamma$ is the constant ratio of specific heats (5/3, typically).

Compared to the more complete MHD equations, the motivation for adopting these ffHD equations lies in getting significant improvement in computational efficiency. 
The full 3D MHD equations encompass eight independent variables, namely density $\rho$, velocities $v_x, v_y, v_z$, magnetic fields $B_x, B_y, B_z$, and thermal pressure $p$, all of which evolve over time.
However, the 3D ffHD equations contain only three independent variables: density $\rho$, field-aligned velocity $v_{\parallel}$, and thermal pressure $p$, and has a background 3D vector magnetic field that does not change in time.
This reduction in time-dependent variables substantially decreases the computational load.
Moreover, the computational time step in solving the ideal 3D MHD set of hyperbolic equations with explicit time-marching schemes is constrained by the Courant-Friedrichs-Lewy (CFL) condition \citep{cour1928}, which relates to the maximum possible speed of information propagation within the computational domain. 
In regions with strong magnetic fields and low density, the Alfvén speed (and the speed of fast magnetosonic waves) can be very high, leading to correspondingly smaller time steps that maintain numerical stability. 
In ffHD, the maximum propagation speed of information is solely dependent on the sound speed, which imposes much less restrictions on the adopted time step. 
This difference between multi-D MHD and ffHD becomes more pronounced in situations with stronger magnetic fields, and ffHD essentially `recovers' the infinite field strength limit of MHD in a computationally efficient manner.

{
However, when non-adiabatic processes are considered, the energy equation for ffHD includes the classic parabolic form of TC:}
\begin{equation}
    \frac{\partial E}{\partial t} + \nabla \cdot \left[\left(E + p\right) v_{\parallel} \bma{\hat{b}}\right] = \rho g_{\parallel} v_{\parallel} + \nabla \cdot \left(q\bma{\hat{b}}\right) - RC + H.
    \label{eq24}
\end{equation}
\begin{equation}
    q = \kappa T^{5/2}(\bma{\hat{b}} \cdot \nabla T).
    \label{eq25}
\end{equation}
{Here, $T$ is temperature, and }$\kappa$ is a constant, in accord with the Spitzer-type heat conductivity \citep{spit1962}.
When $\nabla \cdot \left(q\bma{\hat{b}}\right)$ serves as a source term, an explicit treatment implies a time step $dt_{\mathrm{tc}}$ constrained by the following equation \citep{shar2007, xia2018}:
\begin{equation}
    dt_{\mathrm{tc}} < \frac{\rho f_{tc}}{\left(\gamma-1\right)\kappa T^{5/2}}\min \left( \frac{\Delta x^2}{b_x^2},  \frac{\Delta y^2}{b_y^2},  \frac{\Delta z^2}{b_z^2}\right),
        \label{eq26}
\end{equation}
where $f_{tc} $ is a stability constant which is set to 1/2 for 2D simulations and 1/3 for 3D simulations. 
Note that we assume there is no field-free region when writing Eq.~(\ref{eq26}).

{Despite the computational advantages of ffHD over MHD, the inclusion of parabolic TC introduces significant computational constraints due to the restrictive time step requirements. 
Therefore, addressing the computational bottleneck introduced by the parabolic form of TC is essential for further enhancing the efficiency of the ffHD module.
}

\subsection{Hyperbolic Thermal Conduction}
\label{sec22}
{
As also mentioned in Sect.~\ref{sec1}, to mitigate this bottleneck, we adopt a hyperbolic formulation of TC, following the approach implemented in codes such as MuRAM \citep{remp2017}, Pencil \citep{warn2020}, and MANCHA \citep{nava2022}. 
This hyperbolic approximation transforms the parabolic TC into a hyperbolic form, introducing a finite propagation speed for thermal signals and alleviating the restrictive time step constraints imposed by Equation (\ref{eq26}).
}

This hyperbolic approximation is expressed through an additional hyperbolic equation for the heat flux $q$:
\begin{equation}
    \frac{\partial q}{\partial t} = \frac{1}{\tau} \left( \kappa T^{5/2}(\bma{\hat{b}} \cdot \nabla T) + q\right).
    \label{eq27}
\end{equation}
By this approximation, TC also attains a finite propagation speed, thus the time step is no longer strictly constrained by Eq.~(\ref{eq26}). 
The propagation speed is determined by the value of the parameter $\tau$.
{Its role is to speed up the TC propagation to be comparable with the hydrodynamic part of the simulation.}
If $\tau$ is a very small value, for example, close to the TC timescale $dt_\mathrm{tc}$, the accuracy in treating TC will be very high, approaching that of parabolic TC. 
However, this will result in very short time steps, similar to explicitly solving parabolic TC. 
If $\tau$ is relatively large, the proper treatment of TC may decrease, potentially adjusting thermodynamics at a rate falling behind the evolution speed of the hydrodynamic equations on the left side.

{
In previous works, various expressions of $\tau$ were chosen.
\cite{warn2020} implemented both fixed $\tau$ and adaptive $\tau$ in the Pencil code and conducted tests to compare their performance (see their Table 1). 
Their results indicate that various choices of $\tau$ are valid, but the adaptive $\tau$ is more flexible.
To accommodate simulations of different scales, we choose the adaptive $\tau$ in our implementation within MPI-AMRVAC.
}

Following \cite{remp2017}, in our work, $\tau$ is expressed as:
\begin{equation}
    \tau = \max \left(4 \Delta t, \, \frac{\left(\gamma - 1 \right)\kappa T^{7/2}}{p c^{2}_{s,\mathrm{max}}} \right).
    \label{eq28}
\end{equation}
Here, $c_{s,\mathrm{max}}=\left(\sqrt{\gamma p/\rho}\right)_{\mathrm{max}}$ is the maximum sound speed within the simulation domain, and $\Delta t$ is the time step constrained by the CFL condition of the hydrodynamic part.

\cite{warn2020} provided another expression, and in our ffHD case, the results obtained from the two expressions are similar, differing only by a coefficient, which does not significantly impact the results. 
Additionally, they set an upper limit of 100 s for $\tau$. 
In the simulations we present below, $\tau$ is always much smaller than this upper limit, so we did not intentionally impose this upper limit.
\citet{nava2022}, however, noted that the expression for $\tau$ in Eq.~(\ref{eq28}) did not ensure stability in all their tests. 
Consequently, they adopted a simpler approach by setting $\tau = 4\Delta t$, the lower limit in Eq.~(\ref{eq28}). 
Despite this, we choose to follow the expression from \citet{remp2017} and use Eq.~(\ref{eq28}) in our simulations.

In the next sections, we conduct tests to evaluate the performance of this hyperbolic TC.

\subsection{2D Ring Test}
\label{sec23}
Before applying hyperbolic TC to the complete non-adiabatic 3D MHD simulations, \citet{remp2017} and \citet{nava2022} presented test results for a single scalar model equation of 1D TC using temperature $T$ as a variable:
\begin{eqnarray}
\tau \frac{\partial q}{\partial t} + q &=& \kappa \frac{\partial T}{\partial x},
\label{eq29}\\
\frac{\partial T}{\partial t} &=& \frac{\partial q}{\partial x}.
\label{eq210}
\end{eqnarray}
We conducted similar 1D tests, and our results are consistent with their findings. 
Therefore, we do not display those results here.

Instead, before applying hyperbolic TC in the complete non-adiabatic ffHD system, we conduct a stringent 2D ring test, following the examples set by \citet{parr2005, shar2007, meye2012, xia2018} and \citet{nava2022}.
In this ring test, the continuity and momentum equations are disabled, and in practice the density is fixed to one, while the velocity is fixed to zero, but the frozen magnetic field lines are taken to be circular. 
In the energy equation, all fluxes other than TC are then also disabled, focusing solely on the impact of TC. That means, we are solving the following equation:
\begin{equation}
    \frac{\partial E}{\partial t} - \nabla \cdot \left( q\bma{\hat{b}}\right) = 0,
    \label{eq211}
\end{equation}
where $q$ is calculated from Eq.~(\ref{eq25}) for parabolic and from Eq.~(\ref{eq27}) for hyperbolic TC, respectively.
In this test, thermal conductivity $\kappa T^{5/2}$ is fixed to be 0.01.

The simulation is conducted on a plane extending from $x = -1$ to $x = 1$ and from $y = -1$ to $y = 1$, with a uniform 200$\times$200 grid.
The setup of this test involves the propagation of TC within a 2D annulus.
The annulus is prescribed by setting
\begin{eqnarray}
    b_x &=& \cos \left(\theta + \pi/2 \right),
    \label{eq213}\\
    b_y &=& \sin \left(\theta + \pi/2 \right),
    \label{eq214}
\end{eqnarray}
where $\tan \theta = y/x$.

\begin{figure*}
  \centering
  \includegraphics[width=0.95\textwidth]{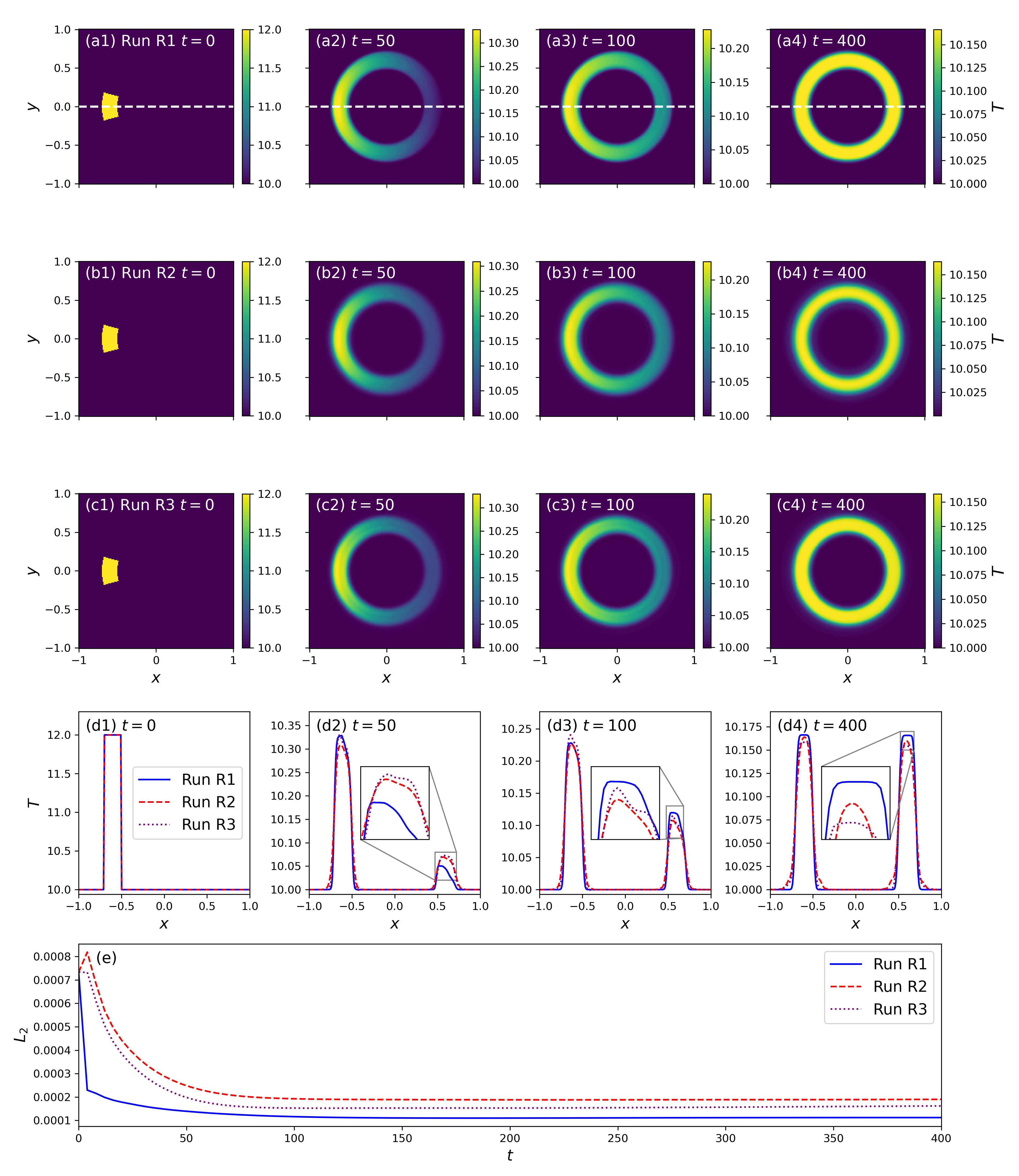}
  \caption{Panels (a1–a4), (b1–b4), and (c1–c4)): Temperature distributions and profiles for the ring test simulations at $t = 0$, 50, 100, and 400 for \texttt{Runs R1}, \texttt{R2}, and \texttt{R3} (see Table~\ref{tb1} for simulation details).
  Panels (d1–-d4): Temperature profiles along the slice indicated in panels (a1–-a4) at  t = 0 , 50, 100, and 400, respectively. The profiles compare \texttt{Runs R1} (blue solid line), \texttt{R2} (red dashed line), and \texttt{R3} (purple dotted line).
  Panel (e): Time evolution of the $L_2$ norm of the temperature difference between the numerical solution and the benchmark solution (\texttt{Run R0}) for \texttt{Runs R1} (blue solid line), \texttt{R2} (red dashed line), and \texttt{R3} (purple dotted line).}
  \label{fig1} 
\end{figure*}

The initial temperature distribution within the annulus is as follows:
\begin{equation}
T=\begin{cases}
     12 & \text{if } 0.5 < \sqrt{x^2+y^2} < 0.7 \text{ and }  \frac{11}{12}\pi < \theta < \frac{13}{12}\pi, \\
     10 & \text{otherwise, }
\end{cases}
\label{eq215}
\end{equation}
and is shown in the first column of Fig.\ref{fig1}. Although our ffHD module focuses on solar prominences, which involve a larger temperature difference, this ring test is a standardized test. 
Previous studies \citep[e.g., ][]{parr2005, shar2007, meye2012, xia2018, nava2022} have used the temperature setting in Eq.~(\ref{eq215}), and we adopted the same setup to ensure consistency and comparability of results.

The first row of Fig.~\ref{fig1} shows the results obtained using the classic parabolic TC, as described by Eq.~(\ref{eq25}), with the STS method.
We refer to this case as \texttt{Run R1}.
Within MPI-AMRVAC, the STS method can be chosen in a Legendre or a Chebyshev variant \citep{pope2021}. 
Here, we present results from the Legendre variant.
Panels (a2)--(a4) display the temperature distribution at $t = 50$ (in dimensionless time units, same below), $t = 100$, and $t = 400$ , respectively. 
It can be observed that the propagation of temperature (or actually the heat flux) is well confined within the prescribed annulus.

The analytical final result should be that the temperature inside the ring is 10.1667, while outside the ring, it remains 10. 
To quantitatively describe the accuracy of this TC, we use the $L_1$, $L_2$ and $L_{\infty}$ norms to measure the error between the final state temperature and the analytical solution, following the approach of previous works. 
The results are presented in Table~\ref{tb1}.
For the classical parabolic TC \texttt{Run R1}, the errors here are consistent with the results from the AMRVAC 2.0 version \citep{xia2018}.

\begin{longtable}{ccccccccc}
\caption{Summary of the ring test}\label{tb1}\\
\hline
Label & Energy Equation & Resolution & Tpye of TC & Flux Scheme & Slope Limiter & $L_1$ & $L_2$ & $L_{\infty}$ \\
\hline
\endfirsthead
\hline
\endfoot
 \texttt{Run R0} & $\frac{\partial E}{\partial t} = \nabla \cdot \left( q\bma{\hat{b}}\right)$ & 1600$^2$ & parabolic & Not Used & Not Used & 0.0001397 & 0.00006451 & 0.09524 \\
 \texttt{Run R1} & $\frac{\partial E}{\partial t} = \nabla \cdot \left( q\bma{\hat{b}}\right)$ & 200$^2$ &parabolic & Not Used & Not Used & 0.005213 & 0.0002535 & 0.08684 \\
 \texttt{Run R2} & $\frac{\partial E}{\partial t} - \nabla \cdot \left( q\bma{\hat{b}}\right) = 0$ &200$^2$ &hyperbolic & TVDLF & MC & 0.01069 & 0.0004818 & 0.08989 \\
 \texttt{Run R3} & $\frac{\partial E}{\partial t} - \nabla \cdot \left( q\bma{\hat{b}}\right) = 0$ &200$^2$ &hyperbolic & TVDLF & WENO-Z & 0.009149 & 0.0003872 & 0.09069 \\
 \texttt{Run RB1} & $\frac{\partial E}{\partial t} = \nabla \cdot \left( q\bma{\hat{b}}\right)$ &200$^2$ &parabolic & Not Used & Not Used & 0.01779 & 0.001543 & 0.1446 \\
 \texttt{Run RB2} & $\frac{\partial E}{\partial t} - \nabla \cdot \left( q\bma{\hat{b}}\right) = 0$ &200$^2$ &hyperbolic & TVDLF & MC & 0.1853 & 0.001609 & 0.1333 \\
\hline
\end{longtable}

Panels (b1)--(b4) display the outcomes from \texttt{Run R2}, which utilizes hyperbolic TC, as per Eq.~(\ref{eq27}). 
From the results, we observe that the dissipation is slightly stronger than with parabolic TC. 
Quantitative results are also shown in Table~\ref{tb1}, indicating that the $L_1$ and $L_2$ norms are approximately twice those of \texttt{Run R1}.
{
Considering that the differences in the temperature distribution plots are relatively small, we extracted a slice along  $y = 0$  to more intuitively display the differences between \texttt{Run R1} and \texttt{R2}. 
In panels (d1)–(d4), we present the temperature profiles along this slice at times  $t = 0$, 50, 100, and 400, using blue solid lines for \texttt{Run R1} and red dashed lines for \texttt{Run R2}. 
At  $t = 50$  (panel d2), we observe that at the right end of the ring (see the zoomed-in region), the temperature in \texttt{Run R2} is slightly higher than that in \texttt{Run R1}. 
This can be interpreted as evidence of the stronger dissipation in hyperbolic TC: the temperature in the perturbed region increases earlier. 
Subsequently, at  t = 100  and 400 (panels d3 and d4), the temperature at the center of the right end of the ring in \texttt{Run R2} becomes lower than in \texttt{Run R1}, while the temperature at the edges of the ring in \texttt{Run R2} remains higher than in \texttt{Run R1}. 
This also reflects the stronger dissipation characteristic of hyperbolic TC.}

{
One contributing factor to the stronger dissipation observed in the hyperbolic TC case is the differing methods of introducing numerical dissipation when calculating parabolic and hyperbolic heat flux. }

Following \citet{shar2007}, \citet{xia2018} treated the parabolic heat flux as a source term on the right-hand side of the equation and introduced a slope limiter when computing $\nabla \cdot \left( q\bma{\hat{b}}\right)$ to suppress numerical oscillations.
This means that, during the ring test, the results are completely unaffected by the use of flux schemes (approximate Riemann solvers) and slope limiters on the left-hand side. 
Conversely, in our implementation, the hyperbolic heat flux is integrated into all other flux components typically present in divergence form on the left side of the energy equation and is computed as part of the overall flux. 
Therefore, numerical dissipation arises not only from the slope limiter but also from the flux schemes.
In this test, we employed the TVDLF flux scheme \citep[Total Variation Diminishing Lax-Friedrichs, see][for details]{kepp2012}. 
To suppress numerical oscillations, this scheme includes a dissipation term proportional to the fastest characteristic speed $c_{\max}$, which is a source of dissipation.
Given that the flux scheme and slope limiters are always activated in practical simulations solving the complete ffHD equations, we have reason to believe that the computational differences between the hyperbolic approximation and the benchmark standard of parabolic TC are acceptably minimal.

Furthermore, since the dissipation is partly caused by the flux scheme and slope limiter, adopting a high-order numerical scheme should mitigate this effect. 
In \texttt{Run R2}, we utilized the second-order monotonized central (MC) limiter \citep{vanl1979}, a robust limiter with relatively low dissipation among second-order total variation diminishing (TVD) limiters. 
{To assess whether a higher-order numerical scheme could reduce the dissipation,} we conducted an additional test by replacing the MC limiter with a fifth-order WENO-Z limiter \citep{borg2008}. 
This run, labeled as \texttt{Run R3}, is presented in panels (c1)--(c4). 
The results are generally consistent with \texttt{Run R2}, with the $L_1$ and $L_2$ norms being slightly smaller.
{The corresponding 1D temperature slices are plotted in panels (d1)–(d4) using a purple dotted line. 
It can be seen that, compared to \texttt{Run R1}, the dissipation was indeed reduced.}

It should be noted that if the numerical scheme described by \citet{kepp2012} is used directly, numerical dissipation can be very significant. 
To mitigate this issue, in \texttt{Runs R2} and \texttt{R3}, we applied modifications using the method described in the appendix of \citet{remp2009} (see their Eq.~(A3) for details; see also \citet{remp2014}), which minimizes numerical dissipation while ensuring numerical stability.
{
Another important consideration is the choice of time step. 
Using a smaller time step can effectively reduce the differences between HTC and PTC, as demonstrated in the 1D tests of \citet{remp2017} and the 2D tests of \citet{nava2022}.
However, in practical simulations, we aim for HTC to have minimal impact on the time step of the ffHD component. 
Therefore, unlike \citet{nava2022}, in this test, the simulation time step is determined by the CFL condition, consistent with our previous ring test using parabolic TC \citep{xia2018}. 
This approach ensures that the time-stepping remains efficient and comparable between HTC and PTC simulations, reflecting realistic conditions in ffHD modeling.
}

{
It should also be noted that in Table~\ref{tb1} and panel (d4) of Fig.~\ref{fig1}, we focus on the errors at the final steady state. 
As shown in the 1D test of \citet{remp2017} and in our results, the hyperbolic TC closely approaches the parabolic TC in the final steady state, showing good agreement with analytical solutions. 
However, hyperbolic TC by design only guarantees convergence to steady states, and the heating and cooling processes in the solar corona are intrinsically very impulsive and dynamic—not steady states. 
Therefore, a more meaningful assessment of the error is to consider the temporal evolution towards reaching the steady-state solution. 
Significant differences may appear during the dynamic process of TC.
}
{
To evaluate whether the error during this evolution process remains within acceptable limits, we calculated the error in the hyperbolic solutions by benchmarking them against the parabolic solutions at each time step. 
Note that in the ring test, only the final steady-state results can be obtained analytically; thus, we do not have the exact temperature distribution at each moment. To address this, we ran a high-resolution (1600$\times$1600) simulation (labeled as \texttt{Run R0}) using parabolic TC for the ring test. We use this result as a benchmark to calculate the $L_1$ and $L_2$ norms at each moment and plotted these errors as a function of time. This approach allows us to assess the accuracy of the hyperbolic TC during the dynamic evolution towards the steady state, which is more relevant for modeling the impulsive and dynamic processes in the solar corona.
}

{
In panel (e), we present the time evolution of the $L_2$ norm for \texttt{Runs R1}, \texttt{R2}, and \texttt{R3} (the evolution trend of the $L_1$ norm is similar, thus not presented here). 
It can be seen that at small times, the errors of hyperbolic TC (\texttt{Runs R2} and \texttt{R3}) are indeed somewhat larger than those of parabolic TC (\texttt{Run R1}), approximately three times larger. 
The errors gradually stabilize at about twice as large later on. 
Overall, they remain within an acceptable range.
}

{
While this ring test in previous works provides valuable insights into the behavior of hyperbolic TC, it does not encompass the large temperature differences associated with solar prominences, nor does it capture the strong non-linearity of TC in the solar corona due to its temperature-dependent conductivity. 
To address these limitations and make our test more applicable to solar coronal conditions, we also conducted a modified ring test incorporating both large temperature contrasts and temperature-dependent thermal conductivity.
}

{
This modified ring test is basically the same as the previous one. 
However, unlike the previous dimensionless runs, this modified test is conducted in a physical domain of [-10, 10]~Mm in both the $x$ and  $y$  directions, representing a typical small-scale region on the Sun. 
The initial thermal spot (i.e., the ring) is defined in the radial range  $5\, \text{Mm} < r < 7\, \text{Mm}$. 
Inside the ring, the temperature is set to  1 MK, while outside the ring, it is  0.01 MK, corresponding to typical coronal and prominence temperatures, respectively. 
Two runs-one with parabolic TC (labeled as \texttt{Run RB1}) and the other with hyperbolic TC (labeled as \texttt{Run RB2})—are conducted, the details of which can be found in Table~\ref{tb1}.
}

\begin{figure*}
  \centering
  \includegraphics[width=0.95\textwidth]{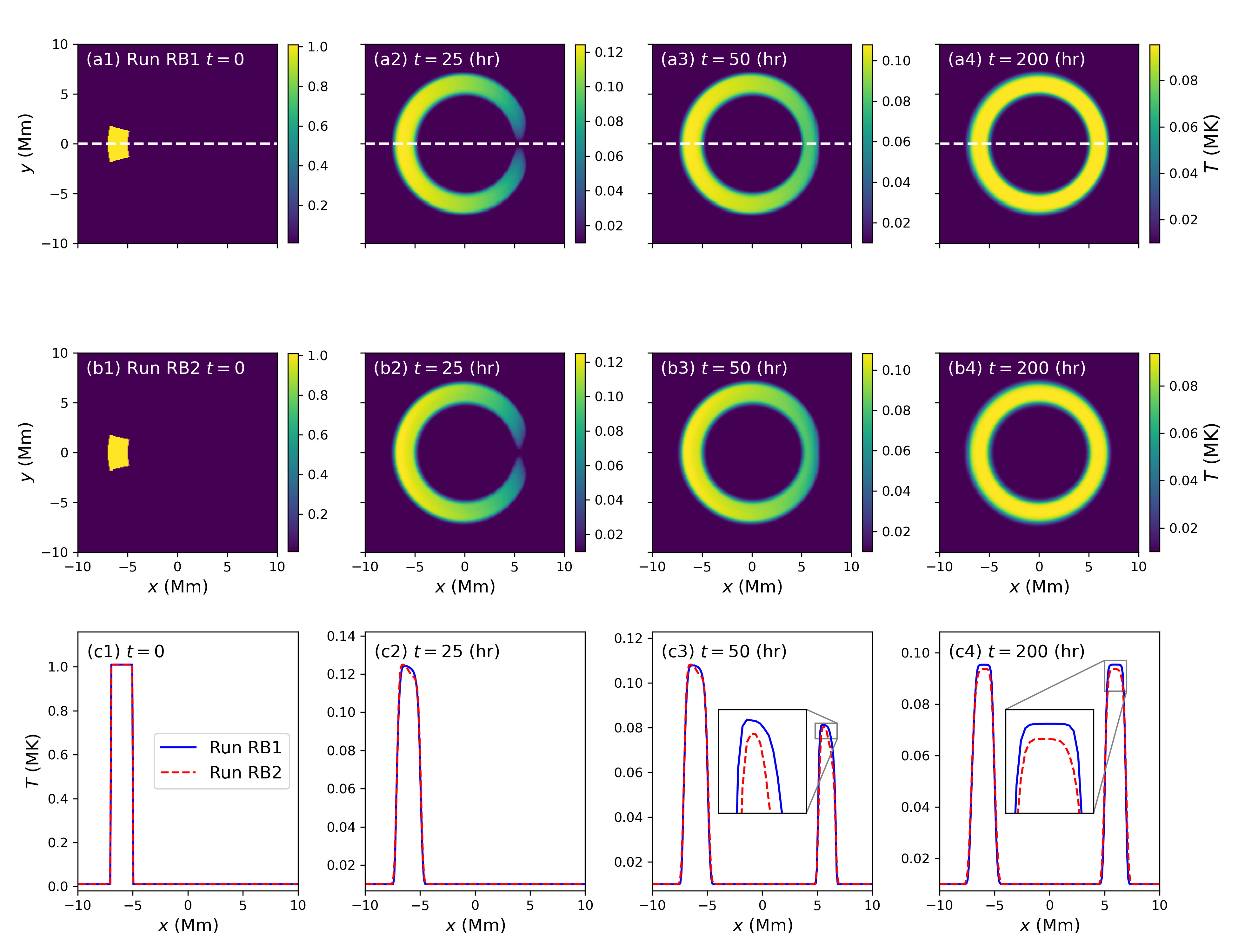}
  \caption{Panels (a1–a4), (b1–b4)): Temperature distributions and profiles for the ring test simulations at $t = 0$, 50, 100, and 400 hr for \texttt{Runs RB1} and \texttt{RB2}).
  Panels (c1–-c4): Temperature profiles along the slice indicated in panels (a1–-a4) at  t = 0 , 50, 100, and 400 hr, respectively. The profiles compare \texttt{Runs RB1} (blue solid line) and \texttt{RB2} (red dashed line)}
  \label{fig1b} 
\end{figure*}

{
The simulation results are presented in Fig.~\ref{fig1b}. 
Similar to Fig.~\ref{fig1}, panels (a1)--(a4) display the temperature distributions of \texttt{Run RB1} at times $t=0$ hr, 25 hr, 50 hr, and 200 hr, respectively, while panels (b1)--(b4) show the temperature distributions of \texttt{Run RB2} at the corresponding times. 
Judging solely from the temperature distribution plots, and much like \texttt{Runs R1} and \texttt{R2}, the differences between \texttt{Runs RB1} and \texttt{RB2} are relatively small, with no significant discrepancies.
}

{
To more clearly illustrate any differences, we extracted a slice along $y=0$ and plotted the temperature distributions of the two runs along this slice at the four time points in panels (c1)--(c4). 
The blue solid line and red dashed line correspond to \texttt{Runs RB1} and \texttt{RB2}, respectively. 
From the results in panels (c3) and (c4), we again observe that, compared to parabolic TC, hyperbolic TC exhibits slightly stronger dissipation, resulting in a slightly lower temperature inside the ring. 
However, overall, the differences are not substantial, and both the absolute and relative errors remain within acceptable ranges.
}

{
The $L_1$ and $L_2$ norms at the final state are also listed in Table~\ref{tb1}. 
It can be seen that, using \texttt{Run RB1} as a reference, the relative errors of \texttt{Run RB2} are similar to those of \texttt{Run R2} relative to \texttt{Run R1}. 
This suggests that the behavior and limitations of hyperbolic TC remain consistent even under solar-like conditions with large temperature differences and temperature-dependent conductivity.
}

\subsection{2D Evaporation--Condensation Test}
\label{sec24}
In this subsection, we apply hyperbolic TC in a more practical solar physics MHD simulation context. 
The model we chose is the 2D evaporation-condensation simulation from \citet{zhou2024}. 
This simulation triggers evaporation by applying artificial localized heating at the footpoints of a 2D potential field and induces condensation in the solar corona. 
Basically, all the settings are the same as described in \citet{zhou2024}, except with a slightly elevated position of localized heating, i.e., specifically \( y_1 \) in Eq. (29) of \citet{zhou2024} is set to 8 Mm. 
This adjustment is made to align with the 3D simulations in the following section. 
Similar to \citet{zhou2024}, we first relax the system for 429 minutes, then introduce artificial localized heating to observe the thermal dynamics induced by evaporation. 

\begin{figure*}
  \centering
  \includegraphics[width=0.95\linewidth]{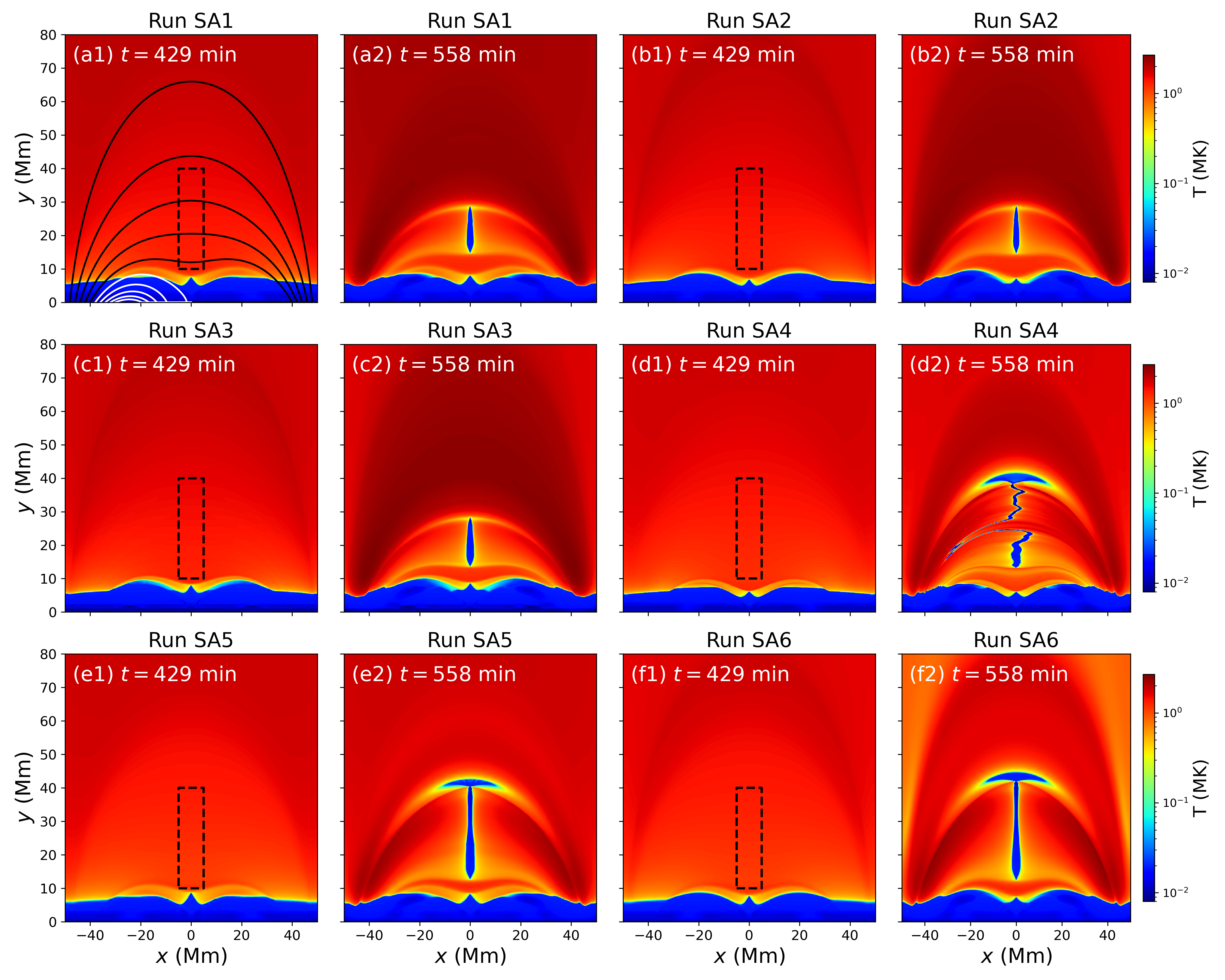}
  \caption{Temperature distribution of the 2D evaporation--condensation test for six different runs (\texttt{Run SA1}, \texttt{SA2}, \texttt{SA3}, \texttt{SA4}, \texttt{SA5}, \texttt{SA6}) at $t=429$ min and $t=558$ min. Black and white solid lines represent the configuration of the magnetic field lines in the top left panel (a1). Dashed lines show the region used for integration in Fig.~\ref{fig3}.}
  \label{fig2}
\end{figure*}

Initially, we conduct the simulation using the ffHD model along with parabolic TC, designated as \texttt{Run SA1}, {which serves as the benchmark solution in this section. 
The comparison between the full MHD and ffHD models was previously presented in \citet{zhou2024}.}
Fig.~\ref{fig2}(a1) displays the temperature distribution at $t = 429$~min, which aligns precisely with Fig.~1(a1) in \citet{zhou2024}. 
The solid lines indicate the configuration of the magnetic field lines. 
Panel (a2) depicts the temperature distribution at the end of the simulation, namely at $t = 558$~min. 
We can clearly see the condensation structure at the center of the simulation domain.
Due to the elevated heating height, the evaporation originates from a position with lower density compared to \citet{zhou2024}, resulting in less material evaporating into the corona, thereby altering the condensation outcome slightly.
Nonetheless, the simulation successfully models the evaporation--condensation process.

Next, we replace the parabolic TC with hyperbolic TC for the same simulation setup, designated as \texttt{Run SA2}. 
Similarly, in panels (b1) and (b2), we showcase the temperature distributions at the end of the relaxation phase ($t = 429$ min) and at the final moment of the simulation ($t = 558$ min), respectively. 
It is apparent that the results obtained with hyperbolic TC are quite similar to those with parabolic TC. 
The location and shape of the condensation structures are analogous.

In the previous ring test, we demonstrated that the computation of hyperbolic TC is dependent on the choice of flux scheme and slope limiter. 
A high-precision numerical scheme can enhance the outcomes of hyperbolic TC. 
In \texttt{Runs SA1} and \texttt{SA2}, we employed the HLL Riemann Solver \citep{hart1983} and the van Leer slope limiter \citep{vanl1974}. 
Next, we substitute the slope limiter with a fifth-order WENO-Z scheme, and this setup is labeled as \texttt{Run SA3}. 
The results are displayed in panels (c1)--(c2).
Again, the simulation results are similar to those of \texttt{Run SA1} and \texttt{Run SA2}, with no significant differences observed.

\begin{figure*}
  \centering
  \includegraphics[width=0.95\linewidth]{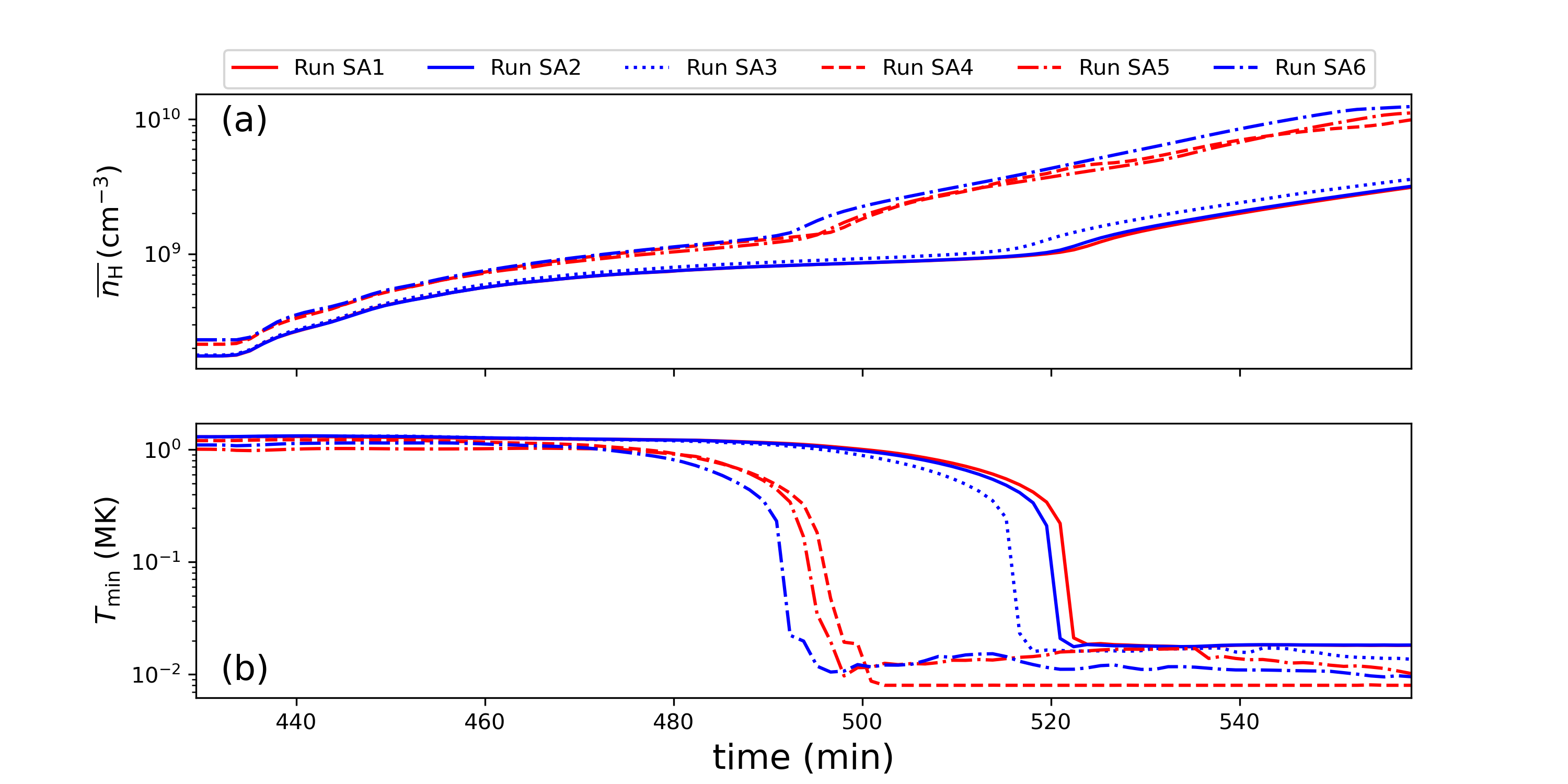}
  \caption{Time evolution of (a) the averaged number density $\bar{n_\mathrm{H}}$ and (b) minimum temperature $T_{\mathrm{min}}$ within the rectangle indicated by the dashed line in Fig.~\ref{fig2} for different runs.}
  \label{fig3}
  
\end{figure*}

To make the comparison quantitative, we select a rectangular region ranging from $x = -5$ to 5 Mm and $y = 10$ to 40 Mm (marked with dashed lines in the $t=429$ panels of Fig.~\ref{fig2}), and track the time evolution of averaged number density of Hydrogen $\overline{n_{\mathrm{H}}}$ within this area. 
The results are displayed in Fig.~\ref{fig3}(a). 
The red and blue solid lines represent \texttt{Runs SA1} and \texttt{Run SA2}, respectively, while the blue dotted line corresponds to \texttt{Run SA3}. 
It is evident that the evolution of $\overline{n_{\mathrm{H}}}$ in \texttt{Runs SA1} and \texttt{Run SA2} almost completely coincide, further indicating that the results of parabolic TC and hyperbolic TC are very close. 
However, $\overline{n_{\mathrm{H}}}$ in \texttt{Run SA3} is slightly higher than in \texttt{SA1} and \texttt{SA2}.
The reason is straightforward.
As mentioned in Sect.~\ref{sec1}, at lower resolutions, the inability to fully resolve the transition region can lead to an underestimation of the upward enthalpy flow, resulting in lower coronal densities. 
The same principle applies here. 
\texttt{Run SA3}, which employs a higher-order limiter, has an effect similar to increasing the simulation's resolution. 
Thus, it can more accurately resolve the transition region and enthalpy flow, and counteract the coronal density underestimated by the low resolution.

Fig.~\ref{fig3}(b) shows the evolution of the minimum temperature $T_{\text{min}}$ within the rectangular area, making it easier to pinpoint the specific moments of condensation. 
Similarly, the evolution patterns in \texttt{Run SA1} and \texttt{Run SA2} are very close, with condensation in \texttt{Run SA2} occurring about 1 minute earlier than in \texttt{Run SA1}. Meanwhile, condensation in \texttt{Run SA3} occurs significantly earlier than in both other runs. 
This is because, generally, higher coronal densities are more conducive to triggering condensation.

According to this explanation, not only utilizing a high-order scheme but also directly enhancing resolution should increase coronal density, thereby accelerating the onset of condensation. 
This has also been validated in the MHD simulations of \citet{zhou2024}. 
Now, in \texttt{Run SA4}, we increase the resolution in \texttt{Run SA1} to assess the impact of higher resolution on ffHD simulations. 
In \texttt{Runs SA1}--\texttt{SA3}, the resolution was the same as in \citet{zhou2024}, which uses a uniform grid with cells of 104 km. 
Here we activate Adaptive Mesh Refinement (AMR), boosting the effective resolution to 13 km. 
The results are displayed in Fig.~\ref{fig2}(d1)--(d2). 
There is a noticeable difference between this result and those of \texttt{Runs SA1$-$SA3}.

We can observe that, in panel (d2), the prominences are longer and the overall condensation area is larger. 
If we define regions colder than 20,000 K as the prominence areas, quantitatively, the lengths of the prominences along the $x=0$ axis in \texttt{Runs SA1$-$SA3} are 13.7, 13.9, and 14.5 Mm, respectively. 
Meanwhile, the prominence length in \texttt{SA4} reaches up to 26.6 Mm, which is nearly twice that of \texttt{Runs SA1--SA3}.
{
Furthermore, we observe that in \texttt{Run SA4}, the prominence exhibits noticeable small-scale variations and significant symmetry breaking that are not present in \texttt{Runs SA1$-$SA3}. 
Although the simulation setup is fully symmetric and, in theory, the results should be perfectly symmetric, numerical simulations inevitably introduce small numerical errors due to finite machine precision and discretization. 
Thermal instability is a highly nonlinear process sensitive to small perturbations; thus, these tiny numerical errors can be amplified over time, leading to asymmetries and the development of small-scale structures in the condensation results. 
In lower-resolution simulations like \texttt{Runs SA1$-$SA3}, numerical diffusion associated with coarser grids can suppress some of these small perturbations, maintaining a more symmetric structure. 
However, at higher resolutions, as in \texttt{Run SA4}, numerical diffusion is reduced, allowing the amplification of small perturbations and resulting in symmetry breaking and more complex prominence morphology. 
Similar observations of symmetry breaking due to the accumulation of numerical errors over time have been reported in previous studies \citep[e.g.,][]{kepp2014}.
}

{
Therefore, it can be seen that high-resolution simulations mainly show three differences: (1) greater amounts of evaporation and condensation, (2) the appearance of small-scale structures, and (3) symmetry breaking. 
Among these, the most physically significant is the first point—the increased evaporation and condensation amounts. 
To compensate for these issues in low-resolution simulations, we have implemented the TRAC method.
}

In recent years, various multi-dimensional TRAC methods have been proposed \citep{zhou2021, iiji2021, john2021}, and we have implemented and validated several variants in MPI-AMRVAC. 
Different TRAC methods may yield slightly different results, but generally, the outcomes are similar. 
Here, we showcase the LTRAC variant from \citet{iiji2021}, primarily because this method better preserves symmetry in the simulations.

\texttt{Run SA5} and \texttt{SA6} incorporate this TRAC correction using parabolic TC and hyperbolic TC, respectively. 
Their temperature distributions are displayed in Fig.~\ref{fig2}(e1)--(e2) and (f1)--(f2) panels.
Indeed, the use of TRAC results in more effective condensation compared to \texttt{Runs SA1$-$SA3}. 
Although our earlier (non-TRAC, but AMR activated) \texttt{Run SA4}, with its higher resolution, shows some symmetry breaking, its prominence morphology remains closer to that observed in \texttt{Run SA5} and \texttt{Run SA6}. 
Quantitatively, the prominence length (along the $x=0$ axis) in panel (e2) is 27.1 Mm, and in panel (f2), it is 27.3 Mm, both closely aligning with the results from \texttt{Run SA4}.

The evolution of $\overline{n_{\mathrm{H}}}$ and $T_{\mathrm{min}}$ over time for \texttt{Runs SA4$-$SA6} is also depicted in Fig.~\ref{fig3}, where the dashed line represents \texttt{Run SA4}, and the red and blue dash-dot lines respectively represent \texttt{Runs SA5} and \texttt{Run SA6}. 
As can be seen, at the end of the relaxation phase, at $t=429$ min, the averaged density (within the rectangle region) $\overline{n_{\mathrm{H}}}$ in these runs is slightly higher than that in \texttt{Runs SA1$-$SA3} (note that the vertical axis is on a logarithmic scale).
Therefore, although the rate of density accumulation is similar (panel (a)), the onset of condensation occurs much earlier (panel (b)), thereby resulting in a final average density difference of approximately four times between \texttt{Runs SA4$-$SA6} and \texttt{Runs SA1$-$SA3}. 
This confirms the trend we previously identified, which suggests that higher densities are more conducive to condensation. 

It is worth mentioning that in this test, the cooling curve used was interpolated from the tables in \citet{colg2008}, as described in \citet{xia2011}. 
In the lower temperature region, below about 18000 K, the radiative losses decrease sharply, approaching zero. 
Thus, the final $T_{\mathrm{min}}$ in \texttt{Run SA1} and \texttt{Run SA2} tends to stabilize around 18000 K. 
However, with higher-order scheme (\texttt{Run SA3}) or with enhanced simulation resolution (\texttt{Run SA4}), lower temperatures can be achieved. 
The introduction of TRAC shows a similar effect.
This is because the stronger dissipation in low-resolution simulations tends to smooth out excessively low temperature structures.

It is worth mentioning that in this test, the cooling curve used was interpolated from the tables in \citet{colg2008}, as described in \citet{xia2011}. 
In the lower temperature region, below about 18,000 K, the radiative losses decrease sharply, approaching zero. 
{
Thus, in \texttt{Runs SA1} and \texttt{SA2}, the final $T_{\mathrm{min}}$ tends to stabilize around 18,000 K due to numerical dissipation inherent in low-resolution or lower-order simulations, which can prevent the temperature from decreasing further. 
However, with a higher-order scheme (\texttt{Run SA3}), enhanced simulation resolution (\texttt{Run SA4}), or the introduction of TRAC (\texttt{Runs SA5} and \texttt{SA6}), lower temperatures can be achieved. 
This is because these approaches allow the simulation to resolve sharper temperature gradients and reach lower temperatures.
}

\begin{longtable}{cccccccc}
\caption{Summary of the 2D evaporation--condensation tests}\label{tb2}\\
\hline
\multirow{2}{*}{Label} & \multirow{2}{*}{Equations} & Type of & \multirow{2}{*}{TRAC} & Flux & Slope & Resolution & Computation \\
& & TC & & Scheme & Limiter & (km) & time (s) \\
\hline
\endfirsthead
\hline
\endfoot
 Run SA0 & MHD & parabolic & Off & HLL & van Leer & 104 & 25151 \\
 Run SA1 & ffHD & parabolic & Off & HLL & van Leer & 104 & 3950 \\
 Run SA2 & ffHD & hyperbolic & Off & HLL & van Leer & 104 & 728 \\
 Run SA3 & ffHD & hyperbolic & Off & HLL & WENO-Z & 104 & 1139 \\
 Run SA4 & ffHD & parabolic & Off & HLL &  van Leer & 13 & 396110 \\
 Run SA5 & ffHD & parabolic & On & HLL & van Leer & 104 & 3723  \\
 Run SA6 & ffHD & hyperbolic & On & HLL & van Leer & 104 & 746 \\
\hline
\end{longtable}

Table~\ref{tb2} gives a summary on all cases from the 2D evaporation--condensation tests, along with the computation times when run in parallel on 4 Intel Xeon Platinum 8468 processor units, each with 48 cores. 
For comparison, we added a test case, \texttt{Run SA0}, which uses the same settings as \texttt{SA1} but is executed under the MHD model. 
In this model, the magnetic field $B_0$ is set to 2 G. 
It is apparent that even with such a modest magnetic field strength, the full MHD simulation consumes significantly more time than the ffHD simulations, with computation times more than 30 times longer compared to the fastest ffHD simulations, \texttt{Run SA2} and \texttt{Run SA6}.

Therefore, compared to the MHD model, using the ffHD model indeed significantly enhances computational efficiency. 
The introduction of hyperbolic TC can further improve computational efficiency. 
Using TRAC correction does not significantly increase the computational load or computation time, and in some cases, as seen in comparisons such as \texttt{Run SA5} versus \texttt{Run SA1}, activating TRAC can even reduce computation time. 
This phenomenon, also observed in \citet{zhou2021}, occurs because changes in the thickness of the transition region can sometimes alter the time step, thereby making simulations with TRAC activated faster.

\section{Applying ffHD Model to the Formation of 3D Prominence}
\label{sec3}
In this section, we will apply the ffHD model to a 3D magnetic flux rope configuration and adopt optimizations introduced in the previous section to simulate the formation and evolution of filaments.

\subsection{Setup of the Simulation}
The setup of the simulation is as follows.
The simulation runs in a 3D box ranging from $x=[-100, 100]$~Mm,  $y=[-75, 75]$~Mm, $z=[0, 100]$~Mm, with a base $80\times60\times40$ grid, and 5 levels of AMR, which leads to an effective resolution of 156 km.

The initial temperature distribution follows our previous work \citep{zhou2018}:
\begin{equation}
T(z)=\begin{cases}
    T_{ch}+\left(T_{co}-T_{ch}\right)\left(1+\tanh\left(\left(z-h_{tr}\right)/w_{tr}\right)\right) & \text{if } z \leq h_{tr}, \\
    \left(7F_c\left(z-h_{tr}\right)/\left(2\kappa\right)+T_{co}^{7/2}\right)^{2/7} & \text{if } z>h_{tr}, \\
\end{cases}
\label{eq31}
\end{equation}
where $h_{tr} = 6$~Mm, $w_{tr} = 500$~km, $T_{ch} = 0.013$~MK, $T_{co} = 1$~MK, $F_c = 2\times10^5\unit{erg}\unit{cm}^{-2}\unit{s}^{-1}$.
$\kappa = 8\times10^{-7}\unit{erg}\unit{cm}^{-1}\unit{s}^{-1}\unit{K}^{-1}$ is the Spitzer-type heat conductivity.
This distribution will result in the temperature of the initial atmosphere ranging from a minimum of $T_{ch}$ (at chromosphere) to a maximum of 1.91~MK (at $z=100$~Mm).

The gravitational acceleration $\bma{g}$ as a function of height $z$ is as follows:
\begin{equation}
\bma{g}=-{g_\odot}{r_\odot}^2/{({r_\odot}+z)^2}{\mathbf{\hat e}_z}.
\label{eq32}
\end{equation}
where $\mathbf{\hat e}_z$ is the unit vector for the $z-$direction, ${g_ \odot} = 274 \unit{m}\unit{s^{-2}}$ is the gravitational acceleration at the surface of the Sun, and ${r_\odot}=691 \unit{Mm}$ is the radius of the Sun.

Subsequently, we assume that at $z=20$~Mm, the number density of hydrogen is $5 \times 10^8\unit{cm}^{-3}$, which allows us to derive the distribution of thermal pressure over the whole simulation domain based on hydrostatic equilibrium. 
Similar to the above 2D simulation, the hydrogen to helium ratio in the atmosphere is 10:1. 
However, unlike the fully ionized condition in the 2D cases, this 3D simulation incorporates considerations of partial ionization. 
The treatment of hydrogen ionization follows the approach outlined in the appendix of \citet{zhou2023}; whereas the treatment of helium ionization is similar to the strategy used in \citet{ni2022}, we assume that the helium ionization is a function $f_{\mathrm{ion}}$ that depends only on temperature $T$, whose expression is:
\begin{equation}
f_{\mathrm{ion}}\left(T\right) = \frac{n_{\mathrm{HeI}}}{n_{\mathrm{He}}} = \frac{n_{\mathrm{HeII}}}{n_{\mathrm{HeII}}+n_{\mathrm{HeIII}}} =
\begin{cases}
    133854/(T/1 \unit{K})^{1.35692} & \text{if } T > 6000 \unit{K}, \\
    1 & \text{if } T \leq 6000 \unit{K}, \\
\end{cases}
\label{eq33}
\end{equation}
where $n_{\mathrm{He}}$, $n_{\mathrm{HeI}}$, $n_{\mathrm{HeII}}$, $n_{\mathrm{HeIII}}$ are the number density of total helium, neutral helium, once-ionized helium and twice-ionized helium, respectively.
Based on the temperature and pressure distributions, along with the ionization strategy mentioned above, we can determine the density distribution.

Regarding the magnetic field, we employ the so-called TDm flux rope model \citep{tito2014}.
We place a pair of dipoles at (0, $\pm$30, -45) Mm, which serve as the background field for the TDm model.
The magnetic rope is then wound around a torus.
We place the torus's plane in the $xz-$plane, so that the main axis of the magnetic flux rope aligns with the perpendicular bisector of the dipoles. 
The center of the torus is located at (0, 0, -45) Mm, with its minor radius $a_{tdm} = 22.5$ Mm and major radius $R_{tdm}=80$~Mm.
The magnetic flux of the rope is calculated according to Eq.~(11) from \cite{tito2014}, ensuring force balance between the magnetic rope and the background dipolar field.
The configuration of the flux rope is shown in Fig.~\ref{fig4} by the solid lines.
We also display the distribution of the magnetic field in the $z$-direction at the bottom plane with the black-white colormap. 
Note that $\hat{b}_z$ represents only the magnitude of the $z$-component of the unit magnetic field vector $B/|B|$, and does not directly correspond to the actual strength of the $z$-direction magnetic field $B_z$.

Similar to the 2D simulations, our energy equation includes three non-adiabatic source terms: heating $H$, optically thin radiative cooling $RC$, and TC, which was the focus of discussion in Sect.~\ref{sec2}.
As with the previous 2D simulations, in this simulation, heating $H$ is composed of background heating $H_{\mathrm{bgr}}$ (used to maintain coronal temperature) and localized heating $H_{\mathrm{loc}}$ (used to induce evaporation).
The background heating $H_{\mathrm{bgr}}$ remains the same as in 2D settings, i.e. $H_{\mathrm{bgr}}=H_0\exp{\left(z/\lambda_0\right)}$, where $H_0=10^{-4}\unit{erg}\unit{cm}^{-3}\unit{s}^{-1}$ and $\lambda_0=50\unit{Mm}$.
For the localized heating, we follow the approach of \citet{xia2016}, concentrating the heating near the two footpoints. 
The specific expression is:
\begin{equation}
  H_{\mathrm{loc}} = {H_1}{R_{\mathrm{ramp}}}\exp\left(-\frac{\left(z-z_{\mathrm{loc}}\right)^2}{\lambda_1^2}\right)\left(\exp\left(-\frac{\left(x-x_{\mathrm{loc}}\right)^2}{\sigma^2}-\frac{y^2}{\sigma^2}\right)+\exp\left(-\frac{\left(x+x_{\mathrm{loc}}\right)^2}{\sigma^2}-\frac{y^2}{\sigma^2}\right)\right).
  \label{eq34}
\end{equation}
Here, $R_\mathrm{{ramp}}$ is a time-dependent ramp function making $H_\mathrm{loc}$ increase/decrease gradually, as introduced later.
Other parameters are chosen as $H_1=8 \times 10^{-3}~\unit{erg}\unit{cm^{-3}}\unit{s^{-1}}$, $\lambda _1 = 3$~Mm, $x_{\mathrm{loc}}= 58.09$~Mm, $z_{\mathrm{loc}}= 10$~Mm and $\sigma = 15$~Mm.

For the radiative losses $RC$, we still adopted the cooling curve defined by \citet{colg2008}, but augmented it at low temperatures with data from \citet{dalg1972}. 
This modification prevents the issue observed in previous 2D simulations, where radiative losses for plasma below 18,000 K were nearly zero, providing a finite amount of radiative loss instead. 
For specific details on the integration of these two cooling curves, we refer to \citet{clae2020} and \citet{herm2021}.
For the TC, we compare two runs using parabolic and hyperbolic TC, designated as \texttt{Run PTC} and \texttt{Run HTC}, respectively.
Additionally, the simulations employ the block-based TRAC method \citep{zhou2021, li2022} to correct the enthalpy flux. 

Regarding boundary conditions, the bottom boundary is fixed {to their initial values} for all variables. 
The top and side boundaries have a minimal impact on the simulations, so we simply use {zero-gradient extrapolation for all the variables}.
In terms of numerical schemes, we utilize the HLL scheme; for \texttt{Run PTC}, we apply the van Leer limiter, and for \texttt{Run HTC}, we use the fifth-order WENO-Z scheme.
{The simulations were performed on KU Leuven/UHasselt’s Tier-2 cluster, wICE, using Sapphire Rapids nodes. 
We utilized 10 nodes, each equipped with 2 Intel Xeon Platinum 8468 CPUs with 48 cores each. 
The \texttt{Run HTC} required approximately 78 hours, whereas the \texttt{Run PTC} took approximately 535 hours.
}

\subsection{General Evolution}
The ramp function $R_\mathrm{{ramp}}$ serves as the switch of the localized heating.
It is set to be,
\begin{equation}
R_\mathrm{ramp}(t) =\begin{cases}
0, & \text{if } t \le t_{\mathrm{relax}} \text{ or } t \ge t_{\mathrm{stop}}, \\
(t-t_{\mathrm{relax}})/t_{\mathrm{ramp}},  &  \text{if } t_{\mathrm{relax}} < t < t_{\mathrm{relax}} + t_{\mathrm{ramp}}, \\
1, & \text{if }  t_{\mathrm{relax}} + t_{\mathrm{ramp}} \le t \le t_{\mathrm{stop}} - t_{\mathrm{ramp}}, \\
(t_{\mathrm{stop}}-t)/t_{\mathrm{ramp}}, & \text{if }  t_{\mathrm{stop}} - t_{\mathrm{ramp}} < t < t_{\mathrm{stop}}, \\
\end{cases}
\label{eq41}
\end{equation}
where $t_{\mathrm{relax}} = 143$~min, $t_{\mathrm{ramp}} = 500$~s and $t_{\mathrm{stop}} = 572$~min.
That means, we first relax the system without $H_{\mathrm{loc}}$ for 143 minutes.
Subsequently, we apply this localized heating for 429 minutes (between $t_{\mathrm{stop}}$ and $t_{\mathrm{relax}}$).
Afterward, we allow the prominence to relax further until $t=716$~min.
{
This relaxation period serves to observe the post-heating evolution of the prominence.}

\begin{figure*}
  \centering
  \includegraphics[width=0.95\linewidth]{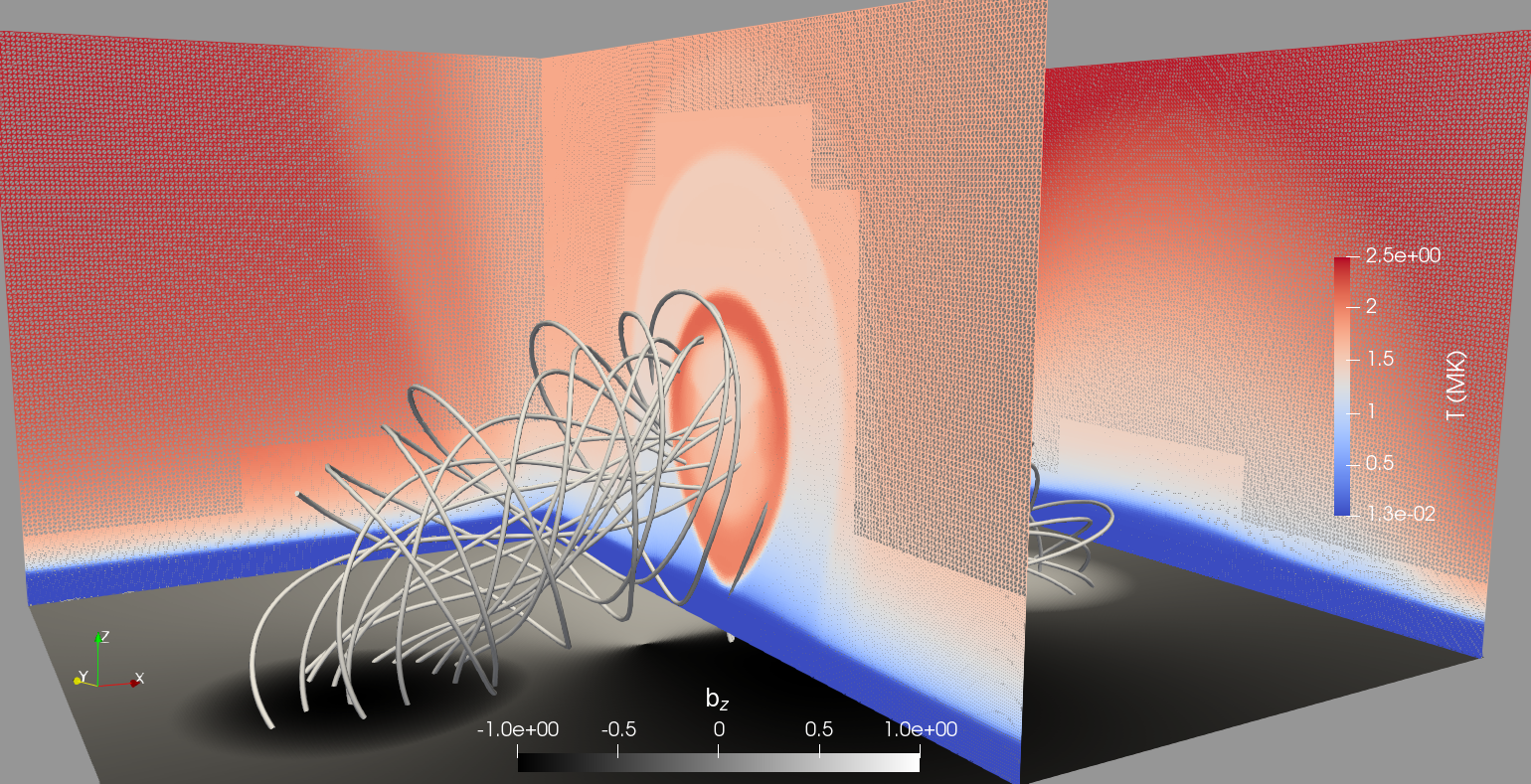}
  \caption{The configuration of magnetic field of the 3D TDm model (solid lines, and the bottom map) and the temperature distribution of some slices (blue-to-red colormap).}
  \label{fig4}
\end{figure*}

The vertical slices in Fig.~\ref{fig4} show the distribution of temperature of \texttt{Run PTC} after relaxation.
\texttt{Run HTC} gives similar results.
We can see that after relaxation, the outline of the magnetic flux rope (referred to as the Quasi-Separatrix Layer, QSL) is heated to higher temperatures. 
Meanwhile, the vicinity of the magnetic flux rope's center and the region outside the flux rope maintain temperatures close to the initial conditions. This specific temperature variation is likely due to the strong twist in the magnetic field lines near the QSL in the TDm model. 
The main axis of our magnetic flux rope follows a circle with a radius of $R = 80$ Mm centered at $z = -45$ Mm, so within the simulation region, the length of the main axis is $2R \cdot \arccos(45.0/80.0) \approx 156$ Mm. 
Due to the inherent twist in the magnetic flux rope (usually 1-2 turns), the lengths of magnetic field lines inside the flux rope are typically longer than the main axis, reaching 1.5--2 times longer, approximately 200--300 Mm. 
However, near the QSL, the flux rope can wind many more turns. 
This is related to the choice of the current density distribution within the flux rope in the TDm model. 
Nevertheless, we believe that the excessive length of these field lines is the reason for their differing dynamics compared to the surrounding areas. 
This is interesting, but since it is not the focus of this paper, we did not investigate it further and leave it for future work. 
Instead, we accept this temperature variation as the realistic distribution obtained after relaxation, and now study the consequences of added localized heating.

\begin{figure*}
  \centering
  \includegraphics[width=0.95\linewidth]{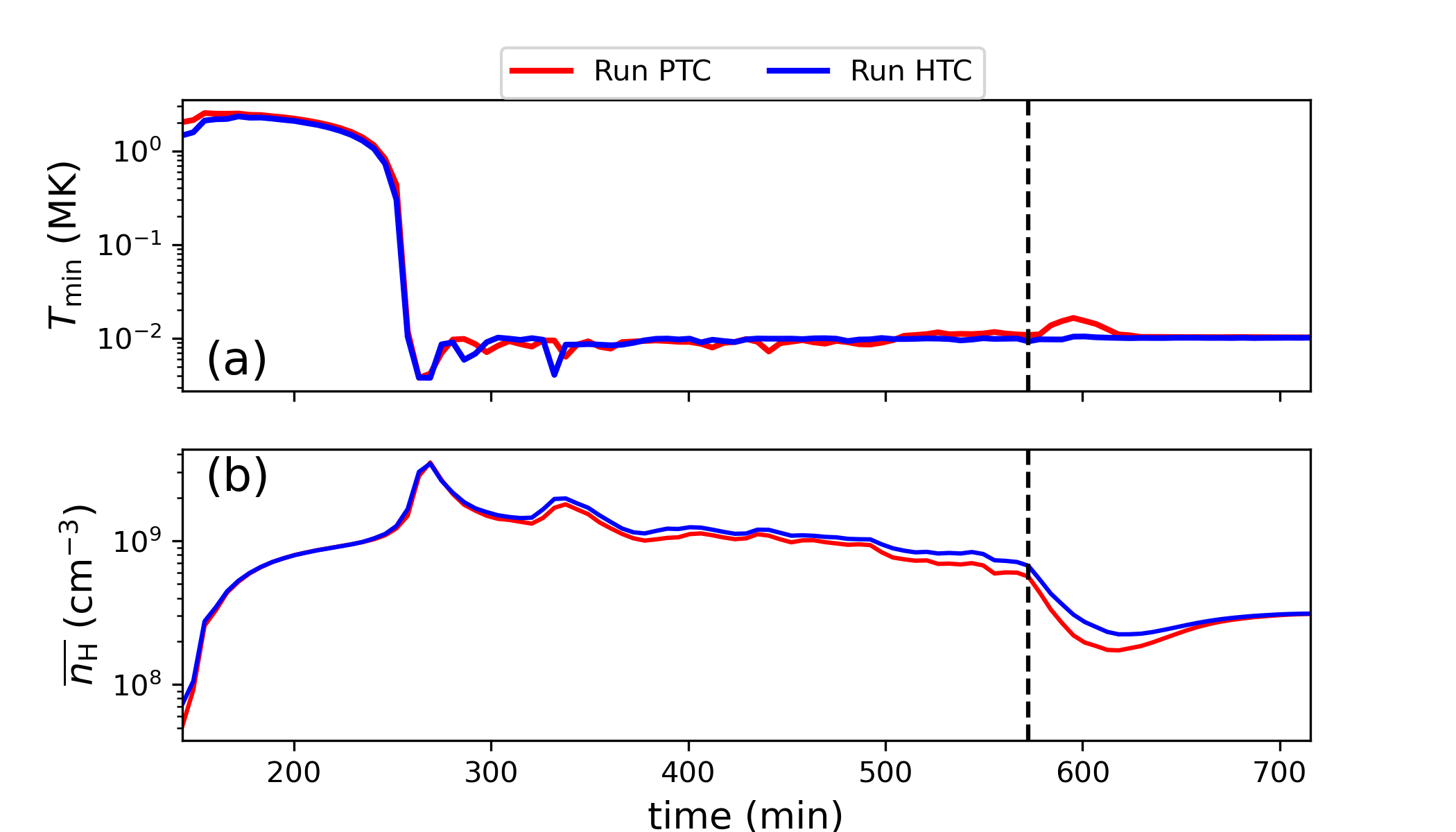}
  \caption{Time evolution of (a) the minimum temperature $T_{\mathrm{min}}$ and (b) the averaged number density $\overline{n_\mathrm{H}}$ within a rectangular volume centered at (0, 0, 32.5) Mm (dimensions: length  $a_{tdm}$ in the $x-$direction), width $a_{tdm}/4$ in the $y-$direction, height $a_{tdm}/2$ in the $z-$direction), for \texttt{Run PTC} (red line) and \texttt{Run HTC} (blue line)}.
  \label{fig5}
\end{figure*}

After initiating localized heating, we first track the evolution of temperature and density at the center of the magnetic flux rope, specifically at the point (0, 0, 35) Mm, using it as a representative example to observe the thermal dynamics. 
The evolution of temperature $T$ at this point is shown in Fig.~\ref{fig5}(a), while panel (b) illustrates the evolution of number density $n_H$ at this point. 
The red line represents the results from \texttt{Run PTC}, and the blue line from \texttt{Run HTC}. 
It is evident that the evolution of both lines approximately coincides.

After initiating localized heating, we first track the evolution of temperature and density {within a representative volume centered around (0, 0, 32.5) Mm, with dimensions corresponding to the minor radius  $a_{tdm}$ in the $x-$direction,  $a_{tdm}/4$ in the $y-$direction, and  $a_{tdm}/2$ in the $z-$direction. 
This position is chosen because the majority of condensation material tends to accumulate in the lower, concave regions of the magnetic flux rope. 
The evolution of minimum temperature $T_\mathrm{min}$ within this volume is shown in Fig.~\ref{fig5}(a), while panel (b) illustrates the evolution of number density $\overline{n_\mathrm{H}}$ within the same region.}
The red line represents the results from \texttt{Run PTC}, and the blue line from \texttt{Run HTC}. 
It is evident that the evolution of both lines approximately coincides.

{
Similar to the trends observed in previous (M)HD simulations \citep{xia2011, xia2016}, initially, there is a rapid increase in $\overline{n_\mathrm{H}}$ accompanied by a slight rise in $T_\mathrm{min}$. 
Subsequently, the rate of increase in $\overline{n_\mathrm{H}}$ slows down, entering a steady rise phase, while $T_\mathrm{min}$ begins to gradually decrease. 
Around $t = 270$ minutes, $T_\mathrm{min}$ begins to drop catastrophically and eventually stabilizes at approximately $10^4$ K, which is typical for prominences. 
Conversely, $\overline{n_\mathrm{H}}$ experiences a rapid increase followed by a gradual decrease, possibly due to the gradual slippage of cold material outside the magnetic dips. 
Nonetheless, $\overline{n_\mathrm{H}}$ remains at a relatively high level, stabilizing at approximately $10^9,\mathrm{cm}^{-3}$.
After turning off the localized heating, $T_{\mathrm{min}}$ remained unchanged, indicating that the main structure of the prominence is stable. 
However, $\overline{n_H}$ decreased. 
This decrease may be due to some material not confined within magnetic dips falling back to the chromosphere \citep{xia2016} and the prominence undergoing a brief expansion following the cessation of siphon flows induced by localized heating \citep{xia2011,zhou2014}.
}

\begin{figure*}
  \centering
  \includegraphics[width=0.95\linewidth]{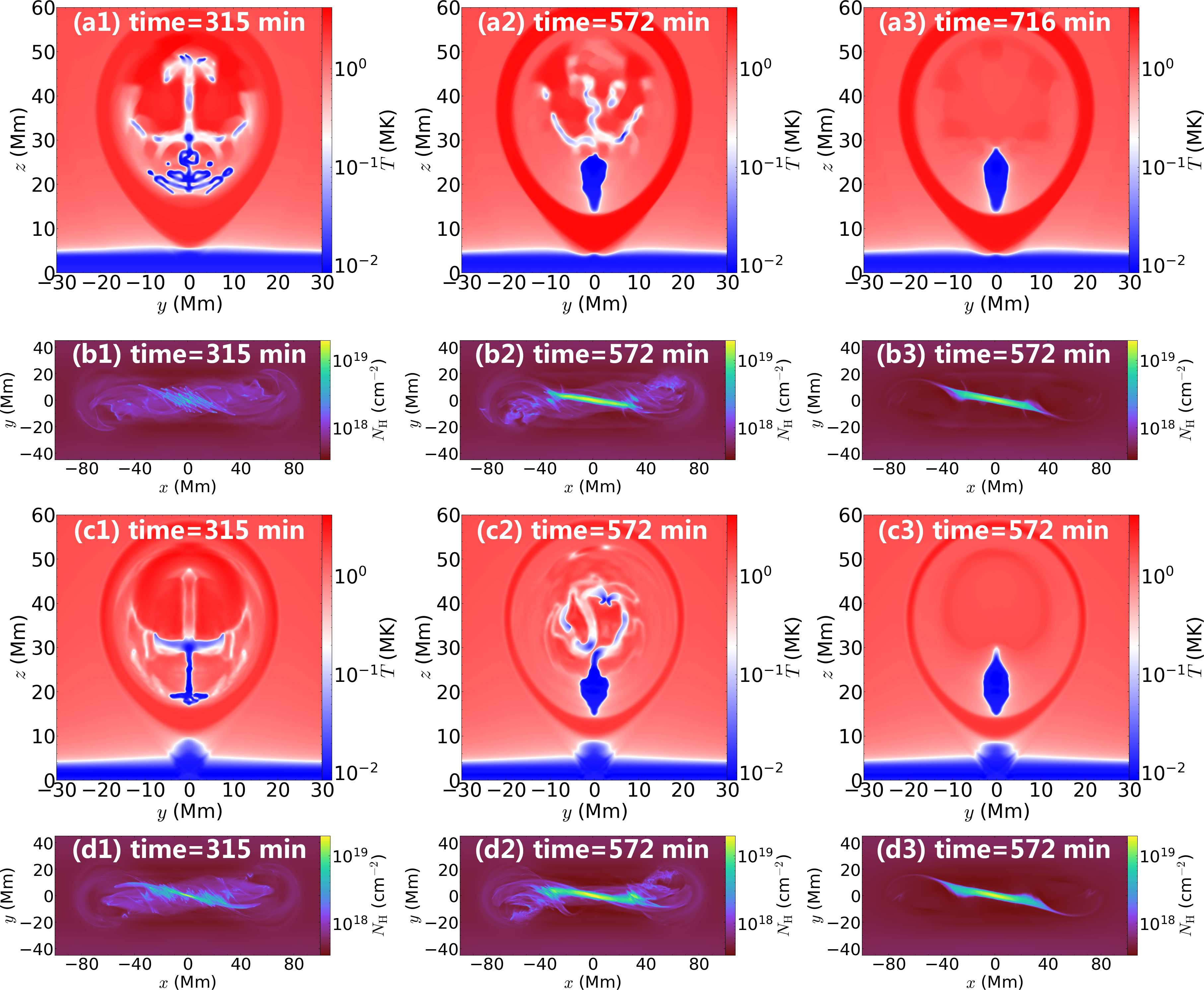}
  \caption{Results of \texttt{Run PTC} and \texttt{Run HTC}: (a1--a3) Temperature distribution of the $x=0$ plane at  $t = 315$,  572 , and  716  minutes, respectively, for \texttt{Run PTC}. (b1--b3) Column density distribution at the same times for \texttt{Run PTC}. (c1--c3) Temperature distribution of the  x=0  plane at the same times for \texttt{Run HTC}. (d1--d3) Column density distribution at the same times for \texttt{Run HTC}.}
  \label{fig6}
\end{figure*}

To provide a more intuitive understanding of condensation within the magnetic dips, we display the temperature distribution on the $x=0$ cross-section at $t=315$, 572, and 716 min during \texttt{Run PTC} in Fig.~\ref{fig6}(a1--a3). 
These times correspond to shortly after condensation formation, after localized heating ends, and at the final moment of the simulation, respectively. 
From panel (a), it is clear that condensation occurs throughout the entire flux rope, primarily concentrating in the lower half where the magnetic field lines with dips are located. 
This concentration becomes more apparent after localized heating ends.
Ultimately, condensation primarily concentrates in a slab-shaped region along the middle of the lower half of the flux rope, akin to prominences in classical 2D models \citep{kipp1957}. 
Simultaneously, only fragmented condensation is present in the upper half of the flux rope, resembling ``coronal rain".
Therefore, we categorize the condensation during the entire process into two parts: one focusing on the ``prominences" in the lower half of the flux rope and another spreading into the ``coronal rain" in the upper half.
After turning off the localized heating and allowing the system to relax for an additional 143 minutes, only the more uniform condensation structures in the lower half of the flux rope remain, while the coronal rain in the upper half dissipates.
Similar results for \texttt{Run HTC} are shown in Fig.~\ref{fig6}(c1–c3).
{
It can be seen that both \texttt{Run PTC} and \texttt{Run HTC} initially experience a dual structure of prominence and coronal rain.
Over time, after the cessation of localized heating, both runs evolve similarly, with only the prominence structures remaining stable.
This indicates that the overall condensation dynamics are consistent between the two methods, despite minor differences in the details of the temperature distribution.}
It is worth noting, as also mentioned in Fig.~\ref{fig4}, that the anomalously high temperatures near the QSL on the periphery of the flux rope are also visible in Figs.~\ref{fig6}. 
However, \texttt{Run PTC} seems to have a thicker high-temperature boundary than \texttt{Run HTC}. 
This is likely because \texttt{Run HTC} used a higher-order scheme, confining the high temperatures to the QSL and preventing dissipation into the surroundings.
 
Fig.~\ref{fig6}(b1--b3) shows the column density $N_\mathrm{H}$ distribution (number density $n_\mathrm{H}$ integrated along the $z-$axis) at the corresponding times in panels (a1--a3). 
Considering that the density of the lower chromosphere is several orders of magnitude higher than that of the corona, including it in the integration would obscure the coronal density distribution. 
Therefore, our integration excludes regions below $z < 7.5$ Mm, focusing only on coronal densities to better represent the distribution of cool material.
These top-down integrated density views show us the following. 
Firstly, at $t = 315$ min (panel (b1)), the column density plot reveals the formation of small, initial high-density condensations. 
These condensations are concentrated in the central part of the magnetic flux rope and seem to align with the local magnetic field lines. 
At $t = 572$ minutes (panel (b2)), a clear formation of the prominence spine is observed. 
Although the temperature plot (panel (a2)) indicates some coronal rain in the upper part of the flux rope, these structures are not visible in the density integration plot, because their integrated densities are much lower than that of the prominence. 
In the final result shown in panel (b3), small structures caused by footpoint evaporation have disappeared, leaving a relatively clean prominence spine. 
Similar results for \texttt{Run HTC} are shown in panels (d1--d3), with an overall evolution consistent with \texttt{Run PTC}, differing only in details.

This elongated prominence showing a clear slab-like configuration and spine structure is different from the more dynamic and fragmented prominences obtained in 3D MHD simulations with similar settings \citep{xia2016}, but it shows some similarity to recent (coronal only) high-resolution prominence formation simulations due to levitation-condensation \citep{jenk2022} or the reconnection-condensation mechanism \citep{donn2024}.

How can we understand these differences? 
Firstly, it is important to note that prominence threads, the building blocks of prominences, are often considered to reflect the direction of the local magnetic field lines \citep{lin2005}. 
The spine of the prominence is typically associated with the polarity inversion line (PIL) in the photosphere. 
Therefore, the threads observed in panel (b1) should basically follow the direction of the local magnetic field lines, forming an angle of about 34 degrees with the $x-$axis. 
However, the prominence spine in panels (b2) and (b3) forms an angle of only 8 degrees with the $x-$axis. 
Similar results have been obtained in pseudo-3D HD simulations using the TDm model \citep{guo2022} (see their Fig. 3).
Thus, the condensation structures on each magnetic field line are actually relatively short. 
Quantitatively, in our TDm model, the lengths of magnetic field lines inside the flux rope vary, but are approximately 200-300 Mm.
However, the lengths of the prominence threads condensed at each magnetic dip are only about 5.7 Mm. 
From the magnetic configuration, these magnetic dips cluster together to outline a narrow filament channel, connecting these threads into a visually continuous prominence spine over 60 Mm long.

This result benefits from the characteristics of the ffHD model, where magnetic field lines are fixed. 
In a full MHD model where magnetic field lines can evolve, the situation would be different. 
In simulations with weaker magnetic fields, the magnetic field lines can be easily dragged by the condensed material. 
For instance, \citet{xia2016} noted that after a magnetic field line forms condensation, it can be dragged and move, causing the cool material to fall to the chromosphere. 
Considering the different geometric parameters and the varying evolution of the evaporation--condensation process on each magnetic field line, the timing of condensation formation will also differ. 
When condensations on some field lines have already fallen to the chromosphere, nearby field lines may still be forming condensations. 
Hence, previous MHD models exhibited fragmented condensation results, unlike the more idealized elongated prominence seen in this work. 
These remaining differences call for further verification through MHD simulations with strong magnetic fields. Note that various important multi-dimensional processes are excluded from the ffHD model by construction: we can have no (magnetic) Rayleigh-Taylor instabilities, no interchange of fieldlines, and certainly no reconnection. 
In many 2D \citep{kepp2014, chan2023, jerc2024} and 3D MHD setups \citep{jenk2022,donn2024}, all these ingredients proved vital in prominence dynamical evolutions. 

We also notice that the PIL (in our case, exactly our $y=0$ axis) and the spine of the resulting prominence in our ffHD model develop an angle. 
In short, this is because in the TDm model, the distribution of magnetic dips is not parallel to the main axis. 
This situation is more noticeable when the minor radius $a$ of the TDm model is larger. 
However, if $a$ is too small, it leads to a higher twist number in the magnetic flux rope, causing kink instability \citep{toro2003} under MHD conditions. 
Therefore, this deviation between PIL and spine might be unavoidable for this particular model setup.

\subsection{Forward Modeling}
\begin{figure*}
  \centering
  \includegraphics[width=0.95\linewidth]{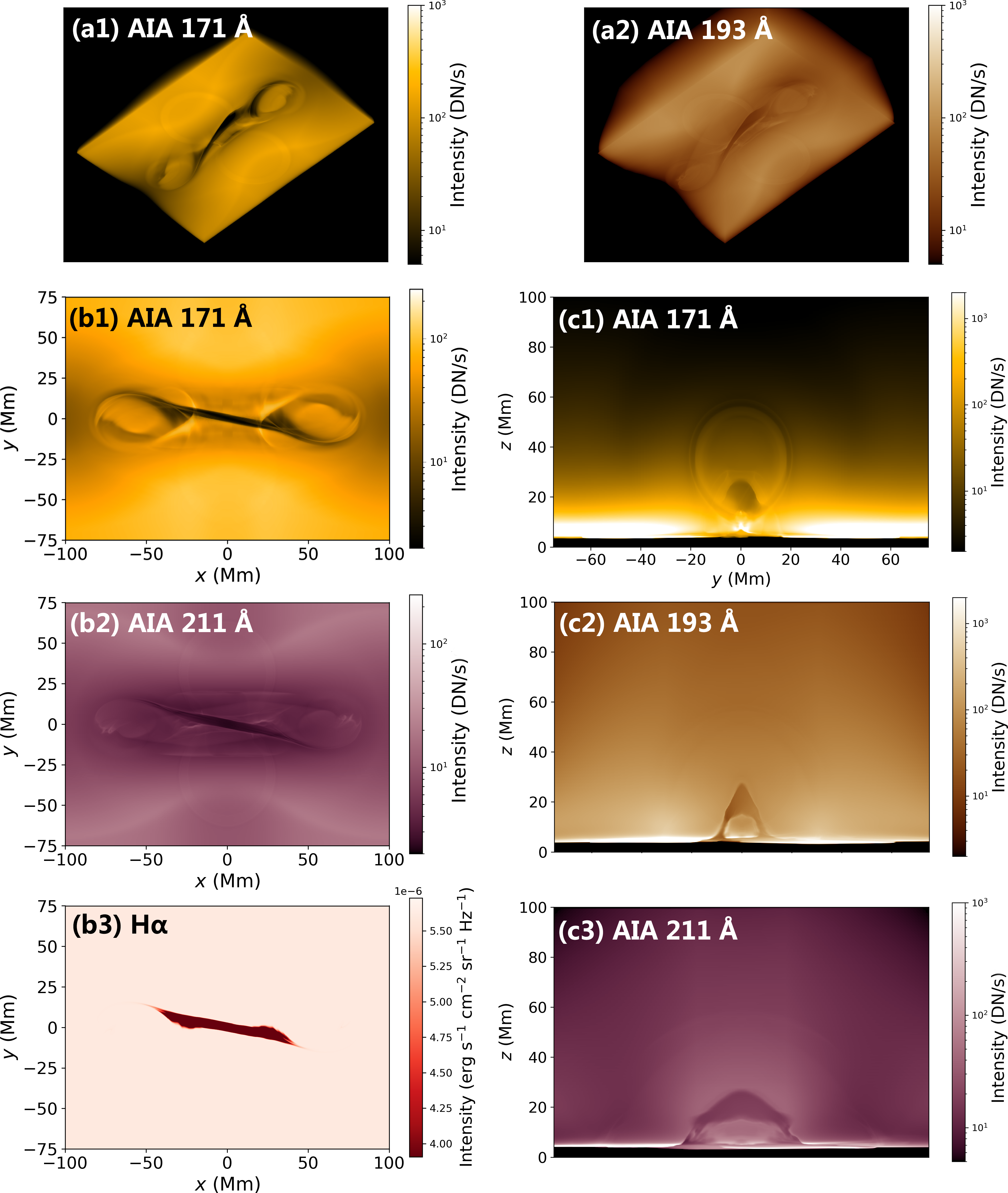}
  \caption{Synthetic images for \texttt{Run HTC}. (a1) Perspective view (at a 45-degree angle to the $x$, $y$, and $z$ axes) in 171 \AA. (a2) Perspective view in 193 \AA. (b1) Top-down view in 171 \AA. (b2) Top down view in 211 \AA. (b3) Top down view in H$\alpha$. (c1) End view (along the $x$-axis) in 171 \AA. (c2) Slightly rotated end view (at a 22.5-degree angle to the $x$-axis) in 193 \AA. (c3) Oblique view (at a 45-degree angle to the $x$ and $y$ axes) in 211 \AA. }
  \label{fig8}
\end{figure*}

To compare the simulated prominence with observations, we synthesized the results into commonly observed wavelengths for prominences, namely the ground-based H$\alpha$ and the 171/211/193 \AA \, bands onboard \textit{SDO}/AIA \citep{leme2012}. 
The EUV bands are nearly optically thin, but our synthesis still considered the absorption caused by photo-ionization in the continuum. 
The specific synthesis methods have been described in \citet{zhou2020} and also referenced in \citet{jenk2022}. 
H$\alpha$ itself is sometimes optically thick and, strictly speaking, requires solving radiative transfer, as demonstrated in recent work \citep{jenk2024} using MPI-AMRVAC and Lightweaver \citep{osbo2021}.
Here, we instead use the approximate method described by \citet{hein2015} and validated for MHD simulations by \cite{jenk2023}. 
Since the approximate method does not account for radiation below the chromosphere/transition region and instead uses the solar surface radiation as the background, we only synthesized the simulation data where $z > 7.5$ Mm.
From the previous analysis, the results of \texttt{Run PTC} and \texttt{Run HTC} are generally similar. 
Therefore, in Fig.~\ref{fig8}, we only show the synthetic images of \texttt{Run HTC} at $t=716$ minutes, the final time of the simulation, as the results of \texttt{Run PTC} are not significantly different.

Although synthetic images during the heating process are also revealing, many previous studies, whether by pseudo-3D \citep{luna2012, guo2022} or through 3D MHD \citep{xia2016}, have already shown and investigated them. 
These works did not present results for the relaxed endstate obtained after the heating is completely turned off. 
Therefore, we focus on this aspect here. 
However, it is important to note that turning off the heating is not essential. 
In reality, it is unlikely that the heating at the footpoints of the prominence (strictly speaking, the magnetic flux rope) would completely disappear. 
The lower atmosphere is always filled with various energetic activities.
Here, turning off the heating allows us to observe if the prominence can remain stable and how much it changes without heating. 


\citet{xia2011, zhou2014} noted that after we turn off the localized heating, the prominence undergoes a rapid outward expansion due to the loss of inward pressure from the evaporation flow. 
\citet{luna2012, guo2022} did not turn off the heating to observe if such expansion occurs. In our simulation, however, we did not observe significant expansion. 
This can be seen from comparing Fig.~\ref{fig6}  panels (b2) and (b3) (or panels (d2) and (d3)). 
In this magnetic flux rope structure, the magnetic dips at the prominence gathering location are deep enough that gravity significantly suppresses any expansion even after losing the pressure from the evaporation flow.

Fig.~\ref{fig8}(a1) and (a2) show {the intensity of AIA} 171 \AA\ and 193 \AA\ from an oblique perspective at a 45-degree angle to the $x$, $y$, and $z$ axes. 
We can clearly see the elongated dark structure roughly along the main axis of the magnetic flux rope, which is the main body of the prominence. 
Additionally, we observe some structures around the footpoints, outlining the magnetic flux rope's contour. 
These structures appear at lower positions (if we only integrate above $z > 7.5$ Mm, these structures disappear). 
The prominence obtained in this simulation is somewhat similar to the results of \citet{donn2024}, forming a more complete overall structure. 
However, there are notable differences: this prominence is a distinctly suspended structure rather than one sliding down to the lower atmosphere. 
This is due to the ffHD model that completely fixes the magnetic flux rope, not allowing any (vertical, cross-field) magnetic slippage.

Panels (b1)--(b3) show top-down views of synthetic images in 171 \AA, 211 \AA, and H$\alpha$ wavebands. 
The shapes in the EUV bands resemble the density integration results in Fig.~\ref{fig6}. 
The 211 \AA\ image is sharper, possibly because 171 \AA\ reveals more lower-layer structures. 
The H$\alpha$ image is very clean, as we only integrated the coronal part without reflecting the lower atmosphere. 
In the H$\alpha$ image, the length of the prominence is about 90 Mm, along with a width of 5.7 Mm, and forms an 8.5-degree angle with the $x-$axis. 
Considering the magnetic flux rope's minor radius is 22.5 Mm, the prominence width is about one-fourth of this. 
This ratio is slightly smaller compared to the results of \citet{guo2022} (see their Fig. 6 and related descriptions). 
Naturally, the relationship between the prominence width and the flux rope width may be influenced by various factors. 
Regarding this point, we are conducting a parameter study for more in-depth research.

Panels (c1)--(c3) show three horizontal views. 
Panel (c1) is an end view of 171 \AA\ along the $x$-axis. 
Besides seeing the suspended prominence structure, we also observe clear loops. 
However, as previously mentioned, these loops are anomalously hot and might not be reliable. 
Our simulation did not exhibit the commonly observed horn structures \citep{Regn2011, Berg2012}, though horn structures are not always present in observations. 
Panel (c2) shows a 193 \AA\ image from a slightly different angle, forming a 22.5-degree angle with the $x-$axis. 
Here, the external contours are barely visible, but more prominence structures are seen, with the main body seemingly connected to the magnetic flux rope footpoints. 
This connection is more clearly shown in the 211 \AA\ image in panel (c3) at a 45-degree angle to the $x-$axis. 
The prominence body is suspended, but some structures near the footpoints extend from the lower atmosphere, possibly remnants of previous evaporation.

These results indicate that the same prominence can appear differently in various wavelengths and viewing angles. 
In numerical simulations with relatively simple setups, this situation is already evident, and in more complex actual observations, the differences in multi-angle views would be even more pronounced.

\section{Summary and Discussion}
\label{sec4}
In modern solar physics research, MHD simulations occupy a crucial position. 
However, even with the rapid growth of computational resources, conducting high-resolution MHD simulations on large scales remains challenging. 
For certain problems where the evolution timescale of the studied phenomenon is shorter than the evolution timescale of the supporting magnetic field, we can simplify the 3D MHD model to essentially 1D HD along field lines, as previously introduced as the ffHD  model in \citet{zhou2024,mok2005}. In our accompanying paper, we derived the ffHD equations and validated their effectiveness in the evaporation--condensation model within a 2D magnetic arcade model.

In this paper, we made further efficiency optimizations to the TC aspect of ffHD, introducing the hyperbolic TC method used in other solar physics software. 
Additionally, we incorporated the TRAC method, which has recently been adopted in (M)HD models \citep{john2019a, zhou2021, iiji2021}, to correct for evaporation in the ffHD model. 
Using these optimizations, we conducted a 3D evaporation--condensation prominence formation simulation based on the TDm model.

The results of this simulation are somewhat similar to those obtained by \citet{guo2022} using a pseudo-3D HD simulation, where 250 magnetic field lines were extracted from a 3D space. 
This similarity is understandable since both models are based on the same assumptions and use the TDm model. 
However, the prominence obtained in this paper is significantly different from that in \citet{xia2016}, which conducted an early prominence-forming 3D MHD simulation with a resolution of 400 km.

In \citet{zhou2024}, we compared the 2D MHD model with the 2D ffHD model based on a magnetic arcade model. 
Under the magnetic arcade model, although the 2D MHD model shows different thermodynamic evolutions depending on the magnetic field strength, qualitatively, its final condensation results are highly similar to those of the ffHD, with only the evolution timescales differing. 
{This indicates that the ffHD model accurately captures the primary condensation processes despite variations in dynamical timescales due to differing magnetic field strengths.}

However, in the more complex 3D magnetic structures, ffHD and MHD exhibit significant differences. 
{Specifically, while MHD simulations (e.g., \citet{xia2016}) produce fragmented condensation structures, our ffHD simulations yield slab-like prominence structures that are more akin to classical models and align closely with many observational results. 
We attribute these differences to the increased complexity of 3D magnetic flux rope configurations and the challenges in maintaining steady-state conditions, which require higher resolution and more robust algorithms. 
Additionally, the ffHD approach perfectly fixes the footpoints of the magnetic flux rope, whereas in actual 3D MHD setups, the details of the implemented boundary treatments (using limited extrapolations from ghost cells to boundary interfaces) may provide additional freedom, potentially affecting the stability and morphology of the condensations.}
Note also that the full MHD scenario has many physical ingredients (e.g., interchange of field lines) that are never incorporated in ffHD runs.

The results obtained here with the ffHD model for a 3D prominence forming by evaporation-condensation are similar to those of some classical models \citep{kipp1957}, as the material eventually collects in (pre-existing) dips. 
Since it is difficult for MHD models to coherently achieve such prominences, in past studies, we artificially inserted prominences at dips in the magnetic flux rope to study their natural oscillations \citep{zhou2018}.
However, not all magnetic dips contain prominence material, and not all prominence material requires dips for support. 
In the ffHD framework, this is a question worth exploring: under different localized heating conditions, what kinds of prominences are formed? 
Especially relevant will be to use spatio-temporally varying heating prescriptions, which have been adopted recently to assess whether prominences develop with a more horizontal or vertical structuring \citep{jerc2024}.

A related issue is the role of background heating. 
Although background heating seems to play a minor role in the evaporation--condensation model, it can significantly affect the final prominence. 
\citet{mok2005, mok2008} were among the first to use the ffHD model to study the impact of coronal background heating on the radiation of active region coronal loops. 
\citet{brug2022} investigated the effect of background heating on prominence formation under 2D MHD conditions. 
However, due to the limitations of the 2D model, they had to introduce a reduction function to describe the impact of different lengths of magnetic field lines in 3D space. 
In 3D simulations, the length of magnetic field lines can be calculated. 
Using the ffHD model to study this issue is advantageous because the magnetic field lines are fixed, and integration only needs to be performed in the initial step, saving subsequent communication time.

Additionally, this study uses the simplest symmetrical heating, resulting in highly symmetrical outcomes without observing significant counterstreaming flows \citep{zirk1998, zhou2020}. 
However, in more complex footpoint heating scenarios, or even with simple asymmetrical heating, the resulting prominences could be completely different. 
This is a topic for further research based on the model presented in this paper.

\section*{Acknowledgment} 
      The authors thank the reviewer for the helpful comments and suggestions.
      YZ acknowledges funding from Research Foundation – Flanders FWO under the project number 1256423N.
      RK and JJ are supported by an FWO project G0B4521N.
      RK, XL and JJ also received funding from the European Research Council (ERC) under the European Union Horizon 2020 research and innovation program (grant agreement No. 833251 PROMINENT ERC-ADG 2018).
      JH is funded by the European Union through the European Research Council (ERC) under the Horizon Europe program (MAGHEAT, grant agreement 101088184). The Institute for Solar Physics is supported by a grant for research infrastructures of national importance from the Swedish Research Council (registration number 2021-00169).
      Visualisations used the open source Python packages \href{https://matplotlib.org/stable/}{Matplotlib} and \href{https://yt-project.org/}{yt}.
      Resources and services used in this work were provided by the VSC (Flemish Supercomputer Center), funded by the Research Foundation - Flanders (FWO) and the Flemish Government.

\vspace{5mm}

\bibliography{main}{}
\bibliographystyle{aasjournal}


\end{document}